\documentclass{article}
\usepackage[a4paper, total={14cm, 23cm}]{geometry}

\usepackage[utf8]{inputenc}

\usepackage{soul} 
\usepackage{todonotes} 
\usepackage{comment} 

\usepackage[bookmarksnumbered = true, colorlinks = true, linktocpage = true, bookmarksopen = true,
    citecolor=blue,%
    filecolor=blue,%
    linkcolor=blue,%
    urlcolor=blue,
	backref=page]{hyperref}
\usepackage[numbers]{natbib} 

\usepackage{graphicx}
\usepackage{float}

\usepackage{amsfonts} 
\usepackage{amsmath}
\usepackage{bm}

\usepackage{tabu}

\usepackage[ruled,vlined]{algorithm2e}

\usepackage{subfiles}

\usepackage{lineno} 

\newcommand{\edits}{\color{black}}

\begin{document}

\begin{titlepage}
    
    \begin{center}
   
    \vspace*{1cm}
    \LARGE
    \textbf{Unlocking ensemble ecosystem modelling for large and complex networks}
       
    \vspace{0.5cm}
    \large{Sarah A. Vollert$^{a,b,*}$, Christopher Drovandi$^{a,b}$, \& Matthew P. Adams$^{a,b,c}$} \\

    \vspace{0.5cm}
    \normalsize
    $^a$Centre for Data Science, Queensland University of Technology, Brisbane, Australia \\

    $^b$School of Mathematical Sciences, Queensland University of Technology, Brisbane, Australia\\

    $^c$School of Chemical Engineering, The University of Queensland, St Lucia, Australia\\

    $^*$Corresponding author. E-mail: sarah.vollert@hdr.qut.edu.au
  
   \end{center}

\vspace{1.5cm}
\begin{center}
    \large
    \textbf{Abstract}
\end{center}
The potential effects of conservation actions on threatened species can be predicted using ensemble ecosystem models {\edits by forecasting populations with and without intervention}. These model ensembles commonly assume stable coexistence of species in the absence of available data. However, existing ensemble-generation methods become computationally inefficient as the size of the ecosystem network increases, preventing larger networks from being studied. We present a novel sequential Monte Carlo sampling approach for ensemble generation that is orders of magnitude faster than existing approaches. We demonstrate that the methods produce equivalent parameter inferences, model predictions, and tightly constrained parameter combinations using a novel sensitivity analysis method. For one case study, we demonstrate a speed-up from 108 days to 6 hours, while maintaining equivalent ensembles. {\edits Additionally, we demonstrate how to identify the parameter combinations that strongly drive feasibility and stability, drawing ecological insight from the ensembles. }Now, for the first time, larger and more realistic networks can be practically simulated {\edits and analysed}.

\vspace{1.5cm}
\begin{center}
    \large
    \textbf{Author summary}
\end{center}

Mathematical models can predict the effects of human actions on an ecosystem, but existing methods of generating these models are only practical for small, simple, food webs. We use advanced statistical sampling techniques to make the model generation process much faster, whilst giving equivalent predictions. Now, these methods can now be used on large, complicated, ecosystems to better explore the complex species interactions that exist in nature.

\end{titlepage}

\nolinenumbers

\section*{Introduction}
Conservation actions aim to help preserve the populations of threatened species, and more generally maintain the health of an ecosystem. {\edits However, it can be challenging to foresee the effects of an intervention across the whole ecosystem, leaving the potential for unintended consequences \citep{prior_2018_unintendedConseq,buckley_2014_unintendedConseq,raymond_2011_unintendedConseq,roemer_2002_unintendedConseq}, such as a shift in predation to increased consumption of a species of interest (e.g., Roemer et al., 2002 \cite{roemer_2002_unintended}).}  Quantitative models can provide critical insights for ecosystem management by forecasting species populations into the future, or in response to both anthropogenic and natural perturbations {\edits \citep{adams2020_TS,Possingham_2001,tulloch_2020}}. However, parameterising these models is challenging. 

There is typically limited information about the model parameters prior to any analysis \citep{baker_2019} due to the difficulty, speed, cost, and uncertainty of expert elicitation and field experiments \citep{Dambacher_2003,Kristensen_2019,Geary_2020}. Consequently, estimates of model parameters necessary to simulate the ecosystem are often poorly constrained and subsequently yield inconclusive forecasts \citep{Novak_2011}. 

Since time-series abundance data is often lacking for model calibration \citep{humbert_2009,mcdonald_2010}, parameters can be constrained based on desired \textit{features} of the ecosystem; two common expected features are feasibility (also referred to as coexistence or persistence) and stability \citep{Dougoud_2018}. {\edits Ensemble ecosystem modelling (EEM) -- an extension of qualitative modelling methods \cite{raymond_2011_unintendedConseq,Dambacher_2003,Melbourne_2012_QM} -- is a method used to generate an ensemble of plausible ecosystem models by randomly sampling parameter values and retaining those that yield feasible and stable ecosystems \cite{baker2017_EEM}. Many studies have used similar methods to simulate ecosystem properties such as these and investigate relationships between network structures, interaction strengths, and ecosystem properties \cite{Dougoud_2018,May_1972,Allesina_2015,Grilli_2017_FS,Stone_2018_FS}. While studies investigating ecological theory could benefit from new parameterisation regimes, we focus on EEM because of its suitability in conservation planning under limited information. In practice, EEM} has been used to assess the indirect consequences of species reintroductions \citep{baker2017_EEM, Pesendorfer_2018_egEEM, Peterson_2021_DirkHartog}, invasive species management \citep{Rendall_2021_EEMeg}, habitat restoration \citep{Bode_2017}, population controls such as baiting \citep{Bode_2017}, and assisted migration \citep{Peterson_2021}. 

Predictions from EEM can inform conservation decisions in the all-too-common situation of limited data availability; however, the process of parameterising the ensemble becomes increasingly computationally intensive as the size of the ecosystem network increases. There can be a very low probability of randomly sampling feasible and stable systems \citep{Allesina_2012_stab}; for example, Peterson and Bode \cite{Peterson_2021} reported fewer than $1$ in $1,000,000$ parameter sets were both feasible and stable for an ecosystem of 15 species. These constraints are even less likely to be satisfied for larger and more complex networks \citep{May_1972,landi_2018}. 

Due to the low probability of generating ecosystem models in which all species stably coexist, much theoretical literature, starting with the classic work of May \cite{Dougoud_2018,May_1972,Allesina_2015,landi_2018}, suggests it is unlikely for complex ecosystems to exist in nature, whereas others have recently proposed explanations for why they do exist -- such as natural selection \citep{barbier_2021,Servan_2018}. {\edits In order to explore these ecological theories and to build decision-making tools,} it is beneficial to model feasible and stable ecosystems -- especially in the absence of time-series data.  Yet in practice, this becomes computationally impractical via random sampling as the food web increases in size \citep{Peterson_2021}. 

In this paper, we exploit established efficient parameterisation methods within Bayesian statistics to present and demonstrate a new method for efficiently generating an ensemble of {\edits parameter sets that define} feasible and stable ecosystem models, inspired by sequential Monte Carlo approximate Bayesian computation (SMC-ABC) \citep{Sisson_2007_SMCABC,drovandi_2011_ABC}. Promisingly, when this new method is compared to the original method proposed by Baker \textit{et al}. \citep{baker2017_EEM} -- hereby referred to as \textit{SMC-EEM} and \textit{standard-EEM}, respectively -- the computational efficiency is increased by several orders of magnitude for larger systems, whilst retaining similar predictions. We demonstrate that SMC-EEM, yields consistent ensembles of ecosystem networks to the standard-EEM method using two common comparisons (parameter inferences and model forecasts) as well as via analysis of model sloppiness \citep{Monsalve_2022} -- a novel model analysis tool \citep{Vollert_2022} that has only recently been applied for comparison of model ensembles \citep{botha_2022}. {\edits Additionally, we demonstrate how this analysis of sloppiness could identify the key parameter combinations driving feasibility and stability, drawing ecological insight from the obtained ensembles.} Therefore, the methods presented here unlock the capabilities of ensemble ecosystem models for representing in, and forecasting for, the complex ecosystem networks that exist in nature.

\section*{Methods}

\subsection*{Ecosystem network modelling}
An ecological community of interacting organisms and their physical environment can be represented as an ecosystem network or food web \citep{Ings_2009}. Ecosystem networks represent the interactions between individual species or groups of species (often referred to as nodes), characterising relationships such as predator-prey, host-parasite{\edits, competitive} or mutualist \citep{Ings_2009,Montoya_2006}. An interaction matrix is used to characterise positive and negative interactions between species that represent a beneficial or detrimental effect on the abundance of the affected species \citep{baker_2019}. By characterising the direct effects of one population on another, the indirect effects that propagate through an ecosystem can be understood and modelled \citep{Novak_2016}. These interaction networks have been analysed both qualitatively \citep{Dambacher_2003,Baker_2018_qual,Levins_1974_qual,dambacher_2002_qual} and quantitatively \citep{adams2020_TS,baker_2019,Novak_2011,baker2017_EEM,Ives_2003} in order to forecast ecosystem population trajectories and predict responses to disturbances.

Ecosystems can be quantitatively modelled in many ways -- such as non-parametric methods \citep{bonnaffe_2022}, empirical dynamic modelling \citep{liu_2022,ye_2015} or stochastic autoregressive models \citep{Ives_2003} (see \cite{Geary_2020} for an overview). Here, we focus on the common quantitative approach of using the generalised Lotka-Volterra equations for forecasting change in ecosystem node abundances over time \citep{adams2020_TS,baker_2019,Murray_2002}, 
\begin{equation}
    \frac{\mathrm{d}n_i}{\mathrm{d}t}= \left[ r_i + \sum_{j=1}^{N} \alpha_{i,j} n_j(t) \right] n_i(t), \qquad \forall i=1,\dots,N,
    \label{eq: LV}
\end{equation}
where $n_i(t)$ is the abundance of the $i$th ecosystem node at time $t$, $r_i$ is the growth rate of the $i$th ecosystem node, $N$ is the number of ecosystem nodes being modelled, and $\alpha_{i,j}$ is the per-capita interaction strength characterising the effect of node $j$ on node $i$. 

If there is no known effect of species $j$ on species $i$, the parameter $\alpha_{i,j}=0$. {\edits However, relationships between species can be prescribed via the sign of the interaction strength parameters. For example, a mutualist relationship would require that both $\alpha_{i,j}$ and $\alpha_{j,i}$ are positive.} Hence, connecting these Lotka Volterra equations to an ecosystem network informs ecosystem-specific information about the interaction strength parameters $\alpha_{i,j}$ in the model. {\edits In this work, we limit consideration to identifying suitable parameter values for a known model structure, rather than identifying appropriate model structures or networks. }

The system represented in Equation \eqref{eq: LV} can be equivalently expressed in a vector form as
\begin{equation}
    \frac{\mathrm{d}\mathbf{n}}{\mathrm{d}t}=[\mathbf{r} + \mathbf{A} \mathbf{n}] \circ \mathbf{n},
    \label{eq: matrix LV}
\end{equation}
where $\mathbf{n}=\{n_i:i=1,...,N\}$ is the vector of species abundances, $\mathbf{r}=\{r_i:i=1,...,N\}$ is the vector of species growth rates, $\mathbf{A}=\{\alpha_{i,j}:i,j=1,...,N\}$ is the $N\times N$ interaction matrix of per-capita interaction strengths between ecosystem nodes, and $\circ$ is the Hadamard or element-wise product.

\subsection*{Feasibility and stability constraints}
The EEM method generates an ensemble of plausible {\edits parameter sets for the generalised Lotka-Volterra model} where there is limited data. To do this, it uses two constraints on the behaviour of the whole ecosystem:  feasibility and stability \cite{baker2017_EEM}. 

{\edits Since there cannot be negative populations, a \textit{feasible} ecosystem is one in which equilibrium populations of all species are positive \citep{Grilli_2017_FS}.} This feasibility condition is met if $n_i^*>0$ for all $i$, where $n_i^*$ is the equilibrium population abundance for node $i$, which is the solution to  
\begin{equation}
    \frac{\mathrm{d}n^*_i}{\mathrm{d}t}=r_i n^*_i + n^*_i \sum_{j=1}^{N} \alpha_{i,j} n^*_j = 0, \quad \quad \quad \forall i=1,...,N.
    \label{eq: equilibrium}
\end{equation}
Following Equation (\ref{eq: matrix LV}), this condition can be rewritten conveniently as
\begin{align}
   \mathbf{n}^* &= -\mathbf{A}^{-1}\mathbf{r},
   \label{eq: equilibrium matrix}
\end{align}
where $\mathbf{n}^*$ is the vector of equilibrium population abundances $n^*_i$ for all species. 

A \textit{stable} ecosystem is one which can recover after small perturbations of species abundances away from equilibrium \citep{Stone_2018_FS}. Specifically, local asymptotic stability (Lyapunov stability) requires that the dynamic system returns to the vicinity of the equilibrium point following a perturbation \citep{Grilli_2017_FS}. To determine if the stability constraint is met the Jacobian matrix $J$ must be evaluated at equilibrium $\mathbf{n}^*$, such that 
\begin{equation}
    J_{ij} = \left. \frac{\partial f_i}{\partial n_j}  \right| _{\mathbf{n}=\mathbf{n^*}}
    \label{eq:jacobian}
\end{equation}
is the $(i,j)$th element of the Jacobian matrix $J$, and $f_i$ is the change in abundance for the $i$th node represented by Equation (\ref{eq: LV}). Equation \eqref{eq:jacobian} indicates that the elements of this Jacobian matrix approximate the effect of species $j$ on species $i$ when the system is close to equilibrium \citep{Stone_2018_FS}.  The dynamic system is considered locally asymptotically stable if the real part of all eigenvalues ($\lambda_i$) of the Jacobian matrix $J$ are negative, i.e. $\mathbb{R} \{ \lambda_i\}<0,  \ \forall i=1,...,N$. For the generalised Lotka-Volterra equations, the elements of the Jacobian matrix evaluated at equilibrium can be calculated as
\begin{equation}
    J_{i,j} = \alpha_{i,j} n_i^*.
\end{equation}

\subsection*{Ensemble ecosystem modelling}
\begin{figure}[H]
    \centering
    \includegraphics[width=\textwidth]{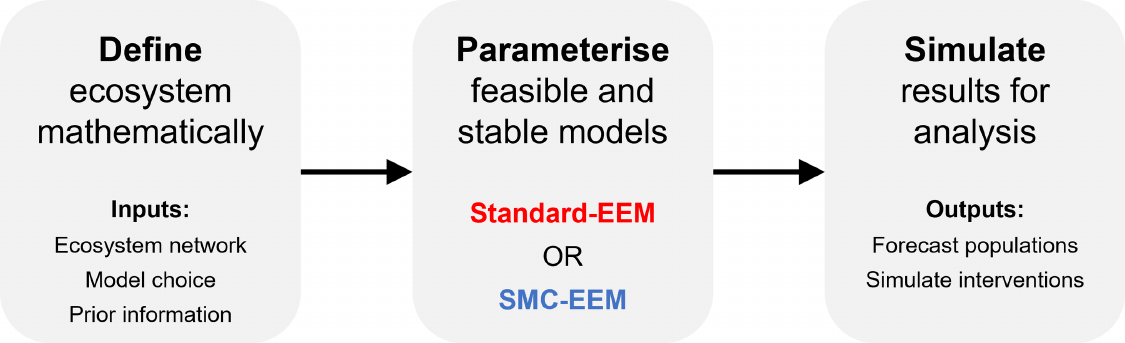}
    \caption{{\bf Overview of the ensemble ecosystem modelling (EEM) process.}
    In the present work, we present the SMC-EEM method and compare it to the standard-EEM method \citep{baker2017_EEM}. The inputs and outputs of the EEM process are the same regardless of the parameterisation method (SMC-EEM or standard-EEM) used.}
    \label{fig:Model_process}
\end{figure}

{\edits Ensemble ecosystem modelling (EEM) aims to produce an ensemble of parameter sets that yield feasible and stable ecosystems for a given ecosystem network structure.} The standard approach to EEM, introduced by Baker \textit{et al}.\ \cite{baker2017_EEM}, is to randomly search a pre-defined parameter space for possible intrinsic growth rate parameters $r_i$ and interaction strengths $\alpha_{i,j}$ that together yield a feasible and stable ecosystem. Specifically, the model parameters $\bm{\theta} \equiv \{\alpha_{ij}, r_i\} _{i,j=1, \dots, N}$ are first sampled from a pre-specified probability distribution which characterises any prior beliefs about the parameter values; this is the prior distribution $\pi(\bm{\theta})$. Next, any sampled parameter sets $\bm{\theta}$ which lead to feasible and stable ecosystems are added to the ensemble of plausible models, creating {\edits an ensemble} of parameter sets {\edits from the target distribution $\pi(\bm{\theta}|\bm{s})$} that have the desired system features $\bm{s}$. Throughout this manuscript, we refer to this random sampling process for generating an ensemble of feasible and stable ecosystems -- described in Algorithm \ref{Alg:Standard-EEM} -- as the \textit{standard}-EEM method. After solving each system of Lotka-Volterra equations, the forecasts are combined to produce an ensemble that can simulate the multitude of potential effects of conservation actions on each of the species within the ecosystem \cite{baker2017_EEM,Pesendorfer_2018_egEEM,Peterson_2021_DirkHartog,Rendall_2021_EEMeg,Bode_2017,Peterson_2021}. A summary of the EEM process is depicted in Fig~\ref{fig:Model_process}.  

\begin{algorithm}[H]
    \While{the ensemble is not sufficiently large}{
    Propose parameter values using any prior beliefs $\bm{\theta}^*\sim \pi(\bm{\theta})$ \\
    \If{the model using $\bm{\theta}^*$ meets the feasibility and stability constraints}{
    Save parameter values $\bm{\theta}^*$ to the ensemble
    }  
    }
    Forecast using the ensemble of ecosystem models \\
    \caption{The standard-EEM algorithm proposed by Baker et al.\ \citep{baker2017_EEM}.}
\label{Alg:Standard-EEM}
\end{algorithm}

While the standard-EEM method can produce a representative {\edits ensemble} of feasible and stable ecosystems, in practice it is too computationally intensive to be practical for large or dense ecosystem networks. We show here that the efficiency of EEM can be greatly improved by exploiting efficient sampling methods developed for Bayesian statistics, such as sequential Monte Carlo-approximate Bayesian computation (SMC-ABC). To explain this, we first demonstrate the connection between EEM and approximate Bayesian computation (ABC).

\subsection*{Approximate Bayesian computation}
ABC is a statistical inference technique used to estimate the parameters of complex models by comparing simulated data to observed data \citep{Beaumont_2019_ABC,Sisson_2018_ABC,sunnaaker_2013_ABC,Beaumont_2010_ABC}. The technique involves simulating data from the model using prior information about the model parameters $\bm{\theta}$ as specified by the prior distribution $\pi(\bm{\theta})$. The simulated data $\hat{\bm{y}}$ (from the model specified by $\bm{\theta}$) is then compared to the observed data ${\bm{y}}$ via a summarisation function $S$ that reduces the full dataset to a set of summary statistics. A discrepancy function $\rho(S(\bm{y}),S(\hat{\bm{y}}))$ is used to measure the similarity between the simulated and observed datasets \citep{sunnaaker_2013_ABC}, and if the simulated data closely matches the observed data, the parameter values are accepted as plausible. The {\edits target (posterior)} distribution, which is a distribution of the parameters conditional on the available data $\pi(\bm{\theta}|\bm{y})$, can then be approximately sampled using ABC accept-reject \citep{beaumont_2002_ABC}, or more efficient methods \citep{Beaumont_2019_ABC} such as Markov chain Monte Carlo ABC (MCMC-ABC) \citep{Martin_2023,gamerman_2006} or {\edits sequential Monte Carlo ABC (SMC-ABC)} \citep{delmoral_2006,chopin_2002}. {\edits For the interested reader, helpful reviews on approximate Bayesian methods can be found in Beaumont et al., \citep{Beaumont_2010_ABC}, Drovandi \citep{drovandi_2017_ABC}, or Sisson et al., \citep{Sisson_2018_ABC}.}

\subsection*{Connections between approximate Bayesian computation and ensemble ecosystem modelling}
While similarities have been drawn between ABC and EEM \citep{baker2017_EEM,Bode_2017}, this connection has not been exploited in the literature to our knowledge. Where ABC uses summary statistics to capture key information in the observed data, EEM applications have no abundance data and instead assume that ecosystems are \textit{observably} feasible and stable. While we suggest that EEM is not an ABC approach in the statistical sense, we propose to frame these system features as summary statistics and adopt ABC-based sampling methods. {\edits In this way, the output of EEM should instead be considered a constraint-informed prior, rather than a posterior distribution -- as feasibility and stability are not directly observed. However, by} placing EEM within an ABC framework, the vast literature on efficient sampling methods developed for ABC can be used to efficiently generate an ensemble of plausible ecosystem networks.

There are many different ABC methods available \citep{Sisson_2018_ABC}, and the simplest is accept-reject ABC. Table \ref{tab:EEM vs ABC AR} reveals that the steps of the standard-EEM method (Algorithm \ref{Alg:Standard-EEM}) are exactly analogous to the ABC accept-reject method \citep{pritchard_1999_ABCrej}. Through both methods, the model parameters $\bm{\theta} = \{ \alpha_{ij}, r_i : i,j=1,...,N\}$ are calibrated using prior information about the parameters and summaries of the data (feasibility and stability).

\begin{table}[!ht]
\begin{tabular}{c|m{0.4\linewidth}|m{0.4\linewidth}} 
    \hline
    \textbf{Step} & \textbf{Standard-EEM method \cite{baker2017_EEM}} & \textbf{ABC accept-reject method \citep{pritchard_1999_ABCrej}}\\ \hline \hline
    1 & \raggedright Generate parameter values $\{\alpha_{ij},~r_i:~i,j~=~1,...,N\}$ independently from distributions. & Draw parameter values from the prior distribution, $\bm{\theta}^* \sim \pi(\bm{\theta})$. \\ \hline
    2 & Calculate equilibrium abundances $n^*_i$ for all species, from the parameter values generated in step 1 (Equation (\ref{eq: equilibrium matrix})), and calculate the eigenvalues of the Jacobian matrix $\lambda_i$ (Equation \eqref{eq:jacobian}). & Simulate data from the model $\bm{\hat{y}}$, using the drawn parameter values $\bm{\theta}^*$.\\ \hline
    3 & Reject model if it is not feasible and stable, i.e.\ if the model does not satisfy $n_i^* > 0$ and $\mathbb{R} \{ \lambda_i\}<0 $ for all $i=1,...,N$. & Reject the parameter values if the discrepancy between modelled and observed data is too large, such that $\bm{\theta}^*$ is rejected if $\rho(S(\bm{\hat{y}}), S(\bm{y})) > \epsilon$ for some tolerance $\epsilon$. \\ \hline
    4 & Repeat steps 1-3 until the ensemble is sufficiently large. & Repeat steps 1-3 until a sufficiently large {\edits ensemble} is obtained.
\end{tabular}
\caption{ {\bf Placing EEM within an ABC framework.} A comparison of the steps in the standard-EEM method and the ABC accept-reject method shows that the two methods are analogous.}
\label{tab:EEM vs ABC AR}
\end{table}

In the ABC accept-reject method depicted in Table \ref{tab:EEM vs ABC AR}, the aim is to minimise the discrepancy ($\rho$) between the modelled and observed data so that they match as much as possible, such that $\rho < \epsilon$ where the target discrepancy $\epsilon$ is small. Equivalently, in the standard-EEM method, the aim is for the features of modelled ecosystems to match what is assumed to be true for a real ecosystem of coexisting species -- feasibility and stability. 

Hence, ABC can be mathematically matched to EEM by introducing a discrepancy function $\rho$ that becomes equal to a target discrepancy of zero ($\epsilon=0$) when the modelled ecosystem is feasible and stable. To this end, we define a discrepancy function $\rho(\bm{\theta})$ for an ecosystem represented by parameters $\bm{\theta}=\{ r_i, \alpha_{i,j}:i,j=1,\dots,N\}$, as 

\begin{align}
\label{eq: discrepancy}
    \rho(\bm{\theta})&= v_f(\bm{\theta}) + v_s(\bm{\theta}),\\
    v_f(\bm{\theta}) &= \sum_{i=1}^{N} \big| \min \{ 0,n^*_i(\bm{\theta}) \} \big|, \label{eq: discrepancy feasibility} \\ 
    v_s(\bm{\theta}) &= \sum_{i=1}^{N} \big| \max\{0,\mathbb{R}\{ \lambda_i\}\} \big|,
    \label{eq: discrepancy end}
\end{align}
where $v_f(\bm{\theta})$ is a measure of infeasibility of all ecosystem nodes (the negativity of equilibrium populations $n^*_i$),  and $v_s(\bm{\theta})$ is a measure of instablility of all ecosystem nodes (the positivity of the real parts of the Jacobian eigenvalues $\lambda_i$). Using the discrepancy function $\rho(\bm{\theta})$ defined in Equations \eqref{eq: discrepancy}-\eqref{eq: discrepancy end}, a feasible and stable ecosystem possesses $\rho(\bm{\theta})=0$; however, any infeasibility or instability will result in $\rho(\bm{\theta})>0$.

\subsection*{Sequential Monte Carlo-approximate Bayesian computation}
By placing EEM within an ABC framework we can take advantage of advanced ABC sampling methods beyond ABC accept-reject sampling. Within the ABC framework, there is a large suite of methods for sampling from the approximate posterior -- such as ABC accept-reject, MCMC-ABC and SMC-ABC \citep{Beaumont_2010_ABC} -- which each present different advantages and disadvantages. In the present work, we used SMC-ABC for sampling because it can be more efficient for applications with a low probability of randomly sampling acceptable parameter values \citep{cerou_2012} and this is the key computational bottleneck in ecosystem generation for large and complex networks. Hence, in the remainder of this section, we provide a brief overview of SMC-ABC as it pertains to ecosystem generation. 

{\edits SMC-ABC works by moving an ensemble of parameter sets through a sequence of distributions, ending at the target distribution \citep{delmoral_2006}. Typically starting with an ensemble drawn from the prior distribution $p_0$, these parameter sets are manipulated to become representative of the next distribution in the sequence $p_1$ and this process is repeated until the ensemble is representative of the target distribution $p_T$. In SMC-ABC, the sequence of distributions $t=0,...,T$ is a sequence of decreasing maximum discrepancies $\epsilon$, such that the $t$th distribution is $p_t(\bm{\theta}|\rho(\bm{\theta}) \leq \epsilon_t)$, where $\epsilon_t \leq \epsilon_{t-1}$. This sequence, whether prespecified or adaptively selected within the algorithm, commonly progresses the ensemble from the prior (maximum discrepancy $\epsilon_0=\infty$) to some target discrepancy (maximum discrepancy $\epsilon_T$).} In this way, SMC-ABC breaks up the sampling problem into a series of simpler problems \citep{DelMoral_2011_SMCABC}. {\edits Provided that the sequence of distributions is chosen sensibly so that the effective sample size throughout the algorithm is maintained at a reasonable level, the sequence itself does not affect the target distribution, merely the speed that the target distribution is obtained.}

{\edits In SMC, a distribution in the sequence is characterised by many independent and weighted parameter sets referred to as `particles'. The weight attributed to each particle is determined by both the prior density and the discrepancy of the parameter set. As such, each particle $\bm{\theta}_i$ contains a proposed value for all model parameters and a weighting, and subsequently an ensemble of $M$ particles make up an empirical approximation of the distribution $p_t$.}

{\edits Each distribution in the sequence, $p_t$, can be approximated by manipulating the ensemble characterising the previous distribution $p_{t-1}$, using importance sampling and MCMC-ABC techniques \citep{drovandi_2011_ABC}. To progress the particles from one distribution to the next, three steps are iteratively applied: reweighting, resampling and moving \citep{Beaumont_2019_ABC,chopin_2002}. }
 
\begin{enumerate}
    \item \textbf{Reweighting}: {\edits The prior density and discrepancy for all particles is calculated and used to weight the particles. This ensures parameter sets that create outputs similar to the observations are more highly weighted. }
    \item \textbf{Resampling}: {\edits Particles are resampled according to their weight, such that high-weighted particles are duplicated and low-weighted particles are eliminated. This focuses the particles into areas of the parameter space that can yield low discrepancies. }
    \item \textbf{Moving}: {\edits MCMC-ABC \citep{gamerman_2006} is used to move the particles according to the current distribution in the sequence $p_t(\bm{\theta}|\rho(\bm{\theta}) \leq \epsilon_t)$. This diversifies the ensemble (avoiding duplicates) by jittering each parameter set relative to its current values. }
\end{enumerate}

{\edits By iterating through these three steps, the cluster of weighted particles can progress through the sequence of distributions to the target distribution.} Algorithm \ref{Alg:SMC-based EEM Overview} shows a summary of an adaptive SMC-ABC method \citep{delmoral_2006}, adapted to the EEM context by building on Drovandi and Pettitt's implementation \cite{drovandi_2011_ABC}. Further details of this algorithm are provided in File S1. 

\begin{algorithm}[H]
    \BlankLine
    \textbf{INITIALISE} \\
    \BlankLine
    Generate {\edits an ensemble} of $M$ particles $\{\bm{\theta}_i\}_{i=1}^{M}$ from the prior distribution, $\pi(\bm{\theta})$ \\    
    \BlankLine
    \BlankLine
    \textbf{REWEIGHT} \\
    \BlankLine
    Evaluate the discrepancy for all particles, $\bm{\rho} = \{\rho(\bm{\theta}_{i})\}_{i=1}^{M}$ \\
    Set the discrepancy threshold $\epsilon_t$ \\
    \BlankLine
    \BlankLine
    \While{there are infeasible or unstable models in the ensemble, $\max(\bm{\rho}) > 0$} {
    \BlankLine
    \textbf{RESAMPLE} \\
    \BlankLine
    Replace {\edits all} particles with a discrepancy greater than the tolerance, {\edits $\rho(\bm{\theta}_i) \geq \epsilon_t$, by duplicating particles with discrepancies below the tolerance} \\
    \BlankLine
    \BlankLine
    \textbf{MOVE} \\
    \BlankLine
    \While{there are many duplicate particles}{
        \For{{\edits each particle that was replaced}}{
        Propose a new set of parameter values $\bm{\theta}^*_i$ from a proposal distribution \\
        Evaluate the discrepancy and prior density, $\rho(\bm{\theta}_i^*)$ and $\pi(\bm{\theta}_i^*)$ \\
        Accept or reject $\bm{\theta}_i^*$ based on a Metropolis-Hastings ratio \\
        }
        \BlankLine
    }
    \BlankLine
    \BlankLine
    \textbf{REWEIGHT} \\
    \BlankLine
    Lower the discrepancy threshold, $\epsilon_t$ \\
    \BlankLine
    }
    \caption{Overview of the SMC-EEM method (see File S1 for full details)}
\label{Alg:SMC-based EEM Overview}
\end{algorithm}

We can think of the ABC accept-reject method (standard-EEM) as ``uninformed'': we reject models that do not fit the constraints, without learning from them. Instead, a more informed sampling method, such as SMC-ABC, utilises information from rejected models. SMC-ABC methods use a sequence of decreasing tolerances, so that parameter values are proposed from an iteratively more ``informed" distribution, rather than the prior \citep{DelMoral_2011_SMCABC}. As a result, SMC-ABC can perform more efficiently than ABC accept-reject for simulating rare events (when the prior and {\edits target distributions} are very different) \citep{Beaumont_2010_ABC}. 

\subsection*{Analysis of model sloppiness}
To compare the ensembles produced by standard-EEM and SMC-EEM, an analysis of model sloppiness can be used. Analysis of model sloppiness is a data-informed sensitivity analysis \citep{Transtrum_2015,brown_2003,gutenkunst_2007} that has recently been shown to provide useful insights for biological and ecological models parameterised using Bayesian inference \citep{Monsalve_2022,Vollert_2022,botha_2022}. In the context of ecosystem generation, analysis of model sloppiness can be used to provide a comparison of the model ensembles generated via different Bayesian methods. 

Whilst {\edits ensembles} can (and should) also be compared based on the estimated marginal parameter distributions, this method can be misleading when individual parameter values are unconstrained. Complementarily, analysis of model sloppiness can be used to compare tightly constrained parameter combinations (e.g.\ products and ratios of parameters) between different {\edits ensembles}, to indicate their similarity even when individual parameter values are relatively unconstrained \citep{botha_2022}. 

{\edits The analysis of model sloppiness uses an eigendecomposition of a parameter-data sensitivity matrix to identify the directions in parameter space, with associated magnitudes, that are most informed by the data \cite{Monsalve_2022,Transtrum_2015}. Here the ``data'' refers to the feasibility and stability constraints.} We use the posterior covariance sensitivity matrix -- the inverse of the empirical covariance matrix of the logarithmically transformed {\edits ensemble} \citep{Monsalve_2022,Vollert_2022} -- to capture how tightly constrained parameters are after parameterisation. {\edits Hence, using this analysis on an ensemble generated via standard-EEM yields the directions in parameter space that are important for obtaining feasible and stable systems. }

{\edits These important directions can be expressed as parameter combinations \citep{Monsalve_2022}, known as eigenparameters $\hat{\theta}_j$:}
\begin{align}
    \hat{\theta}_j &= \theta_1^{v_{j,1}}\theta_2^{v_{j,2}}\cdots\theta_{n_p}^{v_{j,n_p}},
    \label{Eq:Eigenparameter}
\end{align} 
where $\bm{v}_j = [v_{j,1}, v_{j,2}, ..., v_{j,{n_p}}]$ is the $j$th eigenvector of the sensitivity matrix, ${n_p}$ is the number of model parameters, and $\theta_i$ is the $i$th parameter in the model \citep{Vollert_2022,brown_2003}. {\edits Using a logarithmically transformed ensemble allows this eigenparameter to be expressed as a product (as in Eq \eqref{Eq:Eigenparameter}) rather than a sum, which is common in the literature \cite{Monsalve_2022,Vollert_2022,Transtrum_2015}.} Each eigenparameter has a corresponding eigenvalue $\lambda_j$ that indicates how tightly constrained the parameter combination is, such that the largest eigenvalue ($\lambda_1$) corresponds to the most sensitive eigenparameter $\hat{\theta}_1$. {\edits These parameter combinations (expressed as in Equation \eqref{Eq:Eigenparameter}) can be directly analysed to identify important mechanisms \citep{Monsalve_2022}, or visually represented to identify parametric trends \cite{Vollert_2022} that drive the model to match the data (in this case feasibility and stability). For further information about the analysis of model sloppiness method or interpreting eigenparameters, see Monsalve-Bravo et al., 2022 \cite{Monsalve_2022} or Vollert et al., 2023 \citep{Vollert_2022}. }

{\edits Additionally, we can substitute a set of parameter values into each parameter combination. Repeating this process for all parameter sets in an ensemble yields a distribution for each eigenvalue which can be compared to assess ensemble similarity across the important directions in parameter space \citep{botha_2022}. Comparing the distributions of the standard-EEM and SMC-EEM ensembles will reveal whether the important parameter combinations for feasibility and stability are similar between the two methods, indicating ensemble similarity even if individual parameters are unconstrained.}  Hence, the analysis of model sloppiness here provides a critical assessment of the similarity of the ensembles produced by the two different methods of ecosystem network generation (standard-EEM and SMC-EEM). 

\subsection*{Case studies}
The standard-EEM and SMC-EEM methods were compared in two ways. Firstly, the two methods were compared generally across many randomly generated ecosystem network structures (referred to as the ``simulation study''). Secondly, the methods were compared via three case studies representing natural ecosystems. An ecosystem network representing semiarid Australia -- originally used by Baker \textit{et al}.\ \citep{baker2017_EEM} to introduce EEM -- was investigated as an example network where standard-EEM is practical for ecosystem generation within a reasonable computation time. A network of Phillip Island, Australia \citep{Rendall_2021_EEMeg} was used to showcase an example where SMC-EEM is much faster than standard-EEM for ensemble generation. Finally, a coral reef food web network proposed for the Great Barrier Reef \cite{Rogers_2015_reefNetwork} was investigated as an example of interest where the standard-EEM method is computationally impractical. For the simulation study and the three case studies, the computation times and the resulting ensembles produced by each method were compared. 

\subsubsection*{Simulation study}
To generally test the two methods, many ecosystem networks were simulated. Following the practice of May \citep{May_1972} (later replicated by many other studies, e.g., Allesina and Tang \citep{Allesina_2012_stab}), a random matrix theory approach was used, whereby the sign structure of an interaction network was randomly assigned, as follows. 

A network of $S$ species requires a $S \times S$ interaction matrix. The diagonal elements of the matrix (the effect of a species on itself) are negative so that the species populations are self-regulating. {\edits Each off-diagonal element of the matrix was treated independently via a two-step process. Firstly, the interaction was made} non-zero with a probability $c$ -- this connectance parameter specifies the probability of direct interaction between two species \citep{Allesina_2012_stab}. We focused our results on ecosystems generated with a connectance probability of $c=0.5$; however, we also explored varying this probability to $c=0.25$ and $c=0.75$. {\edits Secondly, each non-zero element was allocated either a positive or negative interaction with probability $p=0.5$ (such that there was an equal probability of positive or negative interactions).} Network structures consisting of between 3 and 15 species (inclusive) were generated with this approach.  

For each randomly generated network structure of 3--15 species, 1000 feasible and stable parameterisations were found using the ensemble generation methods discussed previously (standard-EEM and SMC-EEM). We aimed to generate and simulate 1000 ecosystems of each size. However, due to the computational burden of the experiment, we were unable to simulate this many large networks. Instead, there are a minimum of 100 ecosystems simulated for each network size. For each ecosystem network considered in this work (both simulated and natural case studies) the parameterisation used prior distributions of $|\alpha_{i,j}| \sim \mathcal{U}(0,1)$, and $r_i \sim \mathcal{U}(0,5)$ {\edits following Baker et al., 2017} \cite{baker2017_EEM}. 

\subsubsection*{Case study 1: semiarid Australia network}
The two ensemble generation methods (standard-EEM and SMC-EEM) were then applied to an eight-node ecosystem network representing semiarid Australia (see Figure 1b of \cite{baker2017_EEM}). This ecosystem network was previously used to introduce the standard-EEM method and to evaluate the plausible consequences of dingo reintroduction to a national park in Australia \cite{baker2017_EEM}. 

Since standard-EEM has been previously applied to this case study it serves as a useful test case where both methods are expected to generate an ensemble within a practical time frame. In this network, interaction matrix elements that do not represent direct effects of species on each other are set to zero and thus do not require sampling; if this were not the case then ecosystem generation for this (eight-node) network would require sampling of 72 parameters (total 64 interaction matrix elements $\alpha_{i,j}$ and 8 growth rates $r_i$). Instead, this eight-node network has 33 parameters when represented as a generalised Lotka-Volterra model, which is small compared to other ecosystem networks observed in nature that have been quantitatively investigated (e.g.\ Booderee National Park represented as 20 nodes and 163 parameters \citep{baker_2019}).

\subsubsection*{Case study 2: Phillip Island network}
Next, we generated an ensemble of ecosystem models using both standard-EEM and SMC-EEM for a 22 node network which represents Phillip Island, Australia (see Figure 2 of \citep{Rendall_2021_EEMeg}). This network is considerably larger and more complex than the semiarid Australian network -- there are 110 parameters to be estimated when represented as a Lotka-Volterra system -- such that the SMC-EEM method is expected to generate an ensemble faster than the standard-EEM method.

\subsubsection*{Case study 3: Great Barrier Reef network}
Lastly, we demonstrate the benefits of the SMC-EEM method using a case study where it is impractical to use standard-EEM. Rogers \textit{et al}.\ \cite{Rogers_2015_reefNetwork} produced a conceptual 16-node coral reef food web from the literature which depicts a Great Barrier reef ecosystem (see Figure 1 of \cite{Rogers_2015_reefNetwork}). In addition to being a large, this ecosystem network is also densely connected, resulting in an extremely low probability of sampling a feasible and stable model.

\section*{Results}

\subsection*{Simulation study}
\begin{figure}
\centering
\includegraphics[width=0.78\textwidth]{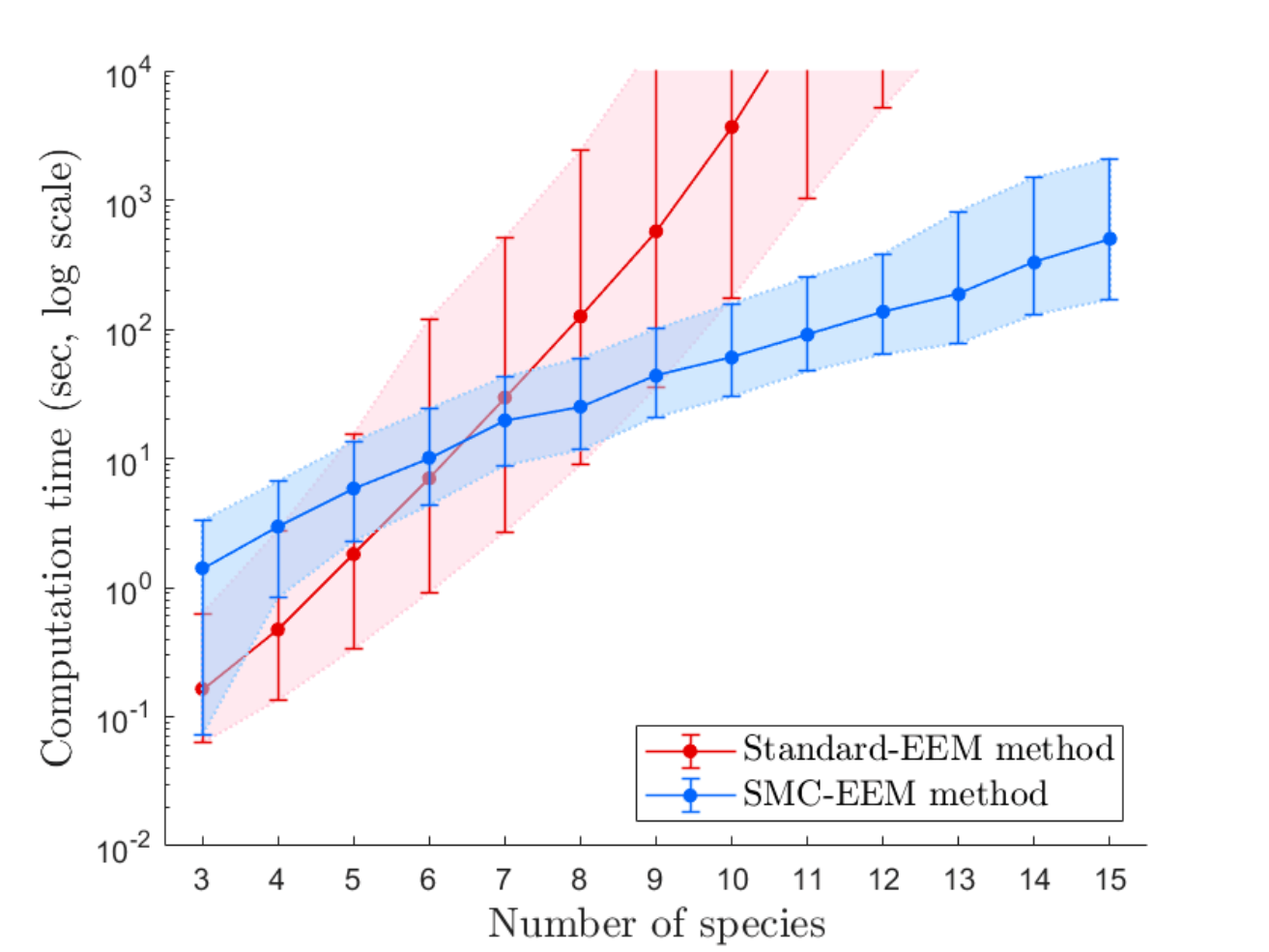}
\caption{{\bf Ensemble generation times for different network sizes.}
The computation time required to parameterise an ensemble of 1000 feasible and stable ecosystem models using both the standard-EEM and SMC-EEM methods. This figure shows the medians (dots) and 7.5--92.5\% quantiles (error bars) of computation times. {\edits Note, the computation time for any one ecosystem network was capped at $10^4$ seconds due to the computational burden of the simulation study.}}
\label{fig:Computation time comparison}
\end{figure}

Our new SMC-EEM method is orders of magnitude faster than the standard-EEM method for larger ecosystems when compared generally across many randomly generated ecosystem network structures (Fig~\ref{fig:Computation time comparison}). We observe that for smaller ecosystems the standard-EEM method may be more computationally efficient due to the additional computational processes required by the SMC-EEM method. This key result also holds for different connectance probabilities $c$ (Fig \ref{Sfig: connectances}). 

\begin{figure}
\centering
\includegraphics[width=\textwidth]{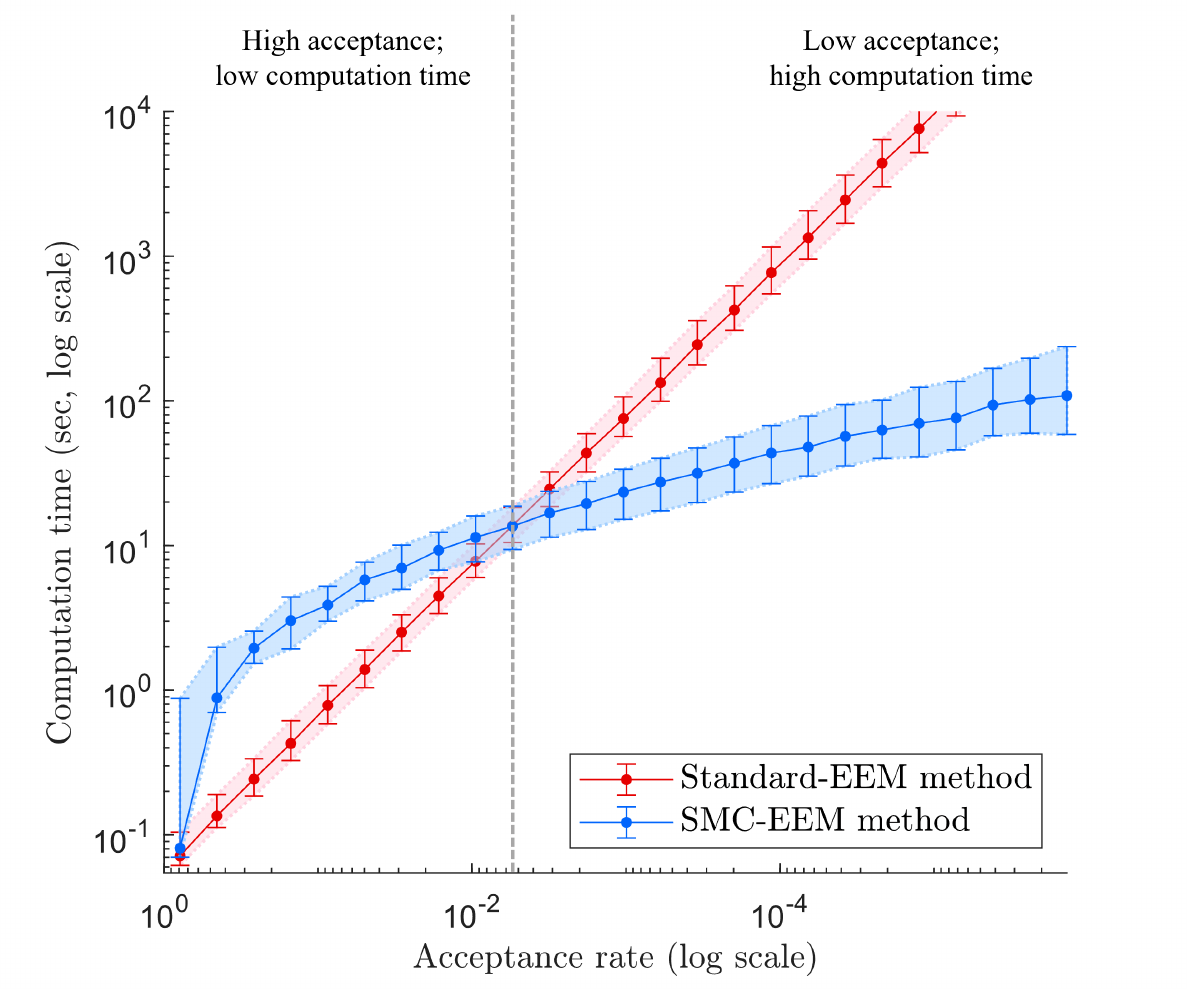}
\caption{{\bf Ensemble generation times for different acceptance rates.}
The parameterisation computation times of Fig~\ref{fig:Computation time comparison} with respect to the \textit{acceptance rate} of the standard-EEM method -- an estimation of the probability of randomly sampling a feasible and stable system given a network with a pre-specified structure. Acceptance rates are logarithmically displayed from 100\% acceptance (left) to very small percentages (right). {\edits Note that the computation time for any one ecosystem network was capped at $10^4$ seconds to maintain practical computations in the simulation study.}}
\label{fig:Computation time acceptance rate}
\end{figure}

More generally, the computation time of the standard-EEM method scales linearly with the probability of randomly selecting parameter values that are feasible and stable (Fig~\ref{fig:Computation time acceptance rate}). This probability -- known as the acceptance rate -- {\edits is an emergent property of the model, prior and constraints, and} can be estimated as the proportion of tested parameter sets that were accepted using standard-EEM. In our simulation study, the SMC-EEM method was computationally more efficient for ecosystems with an estimated acceptance rate smaller than 0.005 (vertical dashed line in Fig~\ref{fig:Computation time acceptance rate}), such that less than 1 in 200 proposed systems are feasible and stable. Here, the SMC-EEM method is faster than the standard-EEM method because fewer parameter values need to be trialled (Fig \ref{Sfig: number of simulations}), making the SMC-EEM method more statistically efficient. {\edits Though standard-EEM can outperform SMC-EEM at high acceptance rates, both methods were computationally inexpensive in these scenarios. In our simulation study, ensembles of 1000 feasible and stable ecosystems could be generated in less than 12 seconds via either method in networks with an acceptance rate greater than 0.005.}

\begin{figure}[H]
\centering
\includegraphics[width=0.95\textwidth]{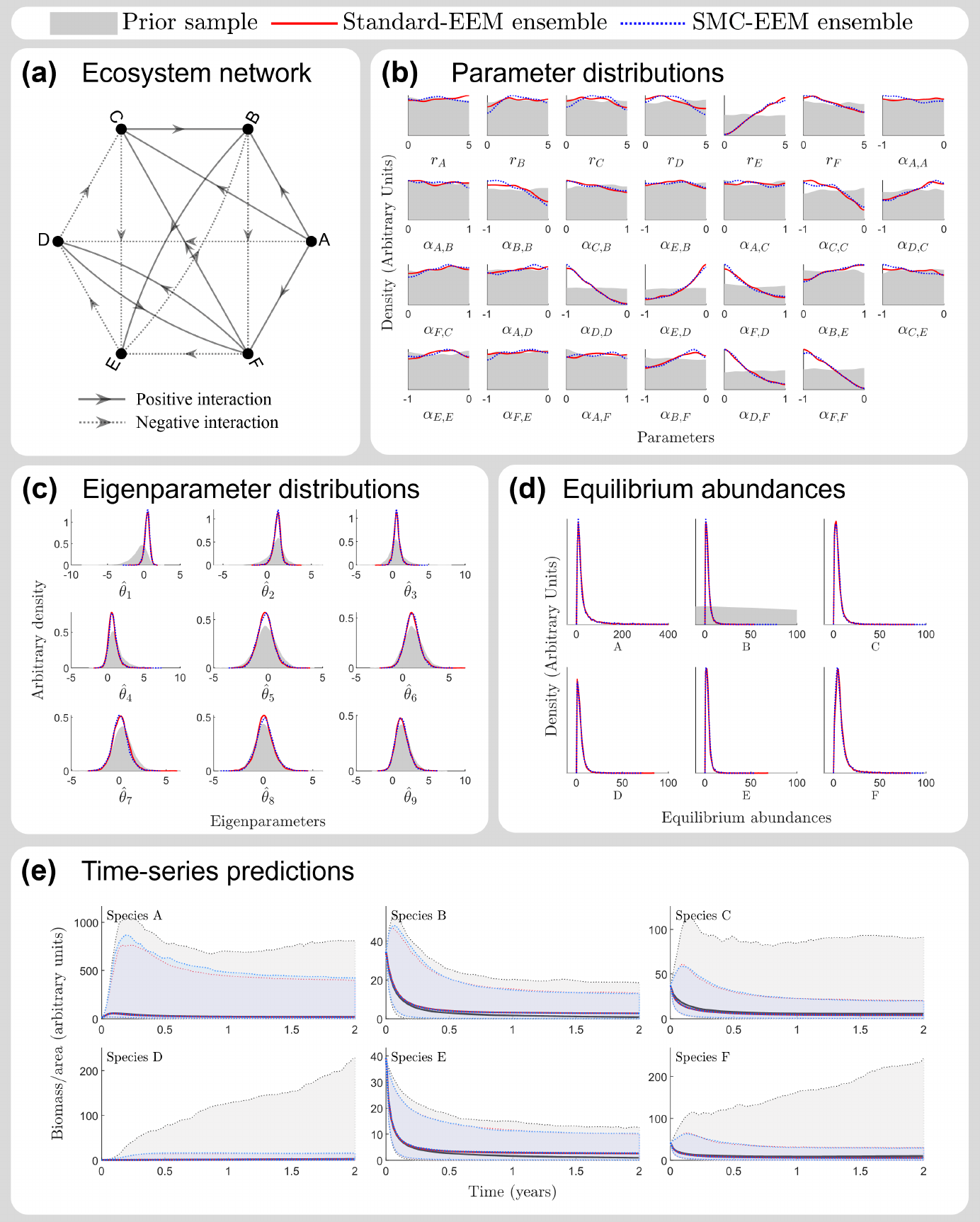}
\caption{{\bf Outputs of a simulated network.}
Example outputs from a randomly chosen ecosystem simulated in Fig~\ref{fig:Computation time acceptance rate} using {\edits ensembles} obtained from the prior distribution (grey), standard-EEM method (red) and SMC-EEM method (blue). In each case, notice that the standard-EEM method and SMC-EEM method produce consistent results that are significantly different to the prior: \textbf{(a)} A six-species ecosystem network generated using $c=0.5$. This example ecosystem has 27 parameters and a 0.037 probability of randomly selecting feasible and stable parameter values. \textbf{(b)} Estimated marginal parameter distributions estimated via both methods and compared to the prior distribution. \textbf{(c)} Marginal distributions of the nine stiffest eigenparameters for each {\edits ensemble} obtained from an analysis of model sloppiness. {\edits \textbf{(d)} The distribution of equilibrium population abundances predicted for each ensemble. Note that the range of equilibrium populations for the prior distribution was $~\mathcal{O}(10^4)$, so is very diffuse (and hence barely visible in these plots) compared to the ensemble-predicted distribution abundances. \textbf{(e)} Time-series predictions of population abundances for each ensemble of ecosystem models using randomly chosen initial conditions (median population prediction and 95\% credible intervals shown).} }
\label{fig:RMT example outputs}
\end{figure}

Additionally, we find that the ensembles of ecosystem models produced by the standard-EEM and SMC-EEM methods are consistent with each other in their estimated parameter distributions, eigenparameter distributions, and time-series predictions (Fig~\ref{fig:RMT example outputs}). For example, for a randomly sampled interaction structure (Fig~\ref{fig:RMT example outputs}a), the SMC-EEM method replicates the outputs of the standard-EEM method in terms of predicted model parameter distributions (blue and red densities in Fig~\ref{fig:RMT example outputs}b). Additionally, from an analysis of model sloppiness, the stiffest eigenparameters (i.e.\ parameter combinations corresponding to the largest eigenvalues of the sensitivity matrix, see Equation \eqref{Eq:Eigenparameter} and surrounding text for more information) also correspond extremely well between the SMC-EEM and standard-EEM methods (blue and red densities in Fig~\ref{fig:RMT example outputs}c). Finally, time-series forecasts of these ecosystems from a common randomly chosen initial condition are virtually indistinguishable between the methods (blue and red shaded regions in Fig~\ref{fig:RMT example outputs}d). 

\subsection*{Case study 1: semiarid Australia network}
\label{Ap: Semiarid results}
For both SMC-EEM and standard-EEM methods, it took less than a minute to generate a 10,000 model ensemble for the semiarid Australia network, though the standard-EEM method was faster (Table \ref{tab:semiarid results}). These computation times are consistent with our previously observed relationship between acceptance rate and computation time (Fig~\ref{fig:Computation time acceptance rate}), as the estimated acceptance rate for this network is $0.11$, which is much larger than $0.005$. As the acceptance rate for the semiarid Australia network is high, only two SMC-ABC iterations were required to generate the ensemble, making the SMC-EEM method statistically inefficient. For this eight-species ecosystem network, with connectance $c=0.39$, the standard-EEM method would be the best choice of method, as it is faster and easier to implement. 

\begin{table}[!ht]
\centering
\begin{tabular}{|l|c|c|}
\hline
& \textbf{Standard-EEM} & \textbf{SMC-EEM}\\ \hline \hline
Computation time (sec) & $5.9$ & $32.4$ \\ \hline
Simulations (number) & $8.7\times 10^4$ &  $10.3\times10^4$\\ \hline
\end{tabular}
\caption{{\bf Computational requirements for the semiarid Australia network.} Computation time and the number of simulations required to generate an ensemble of 10,000 models using both the standard-EEM and SMC-EEM methods for the semiarid Australian ecosystem network.}
\label{tab:semiarid results}
\end{table}

\begin{figure}
\centering
\includegraphics[width=\textwidth]{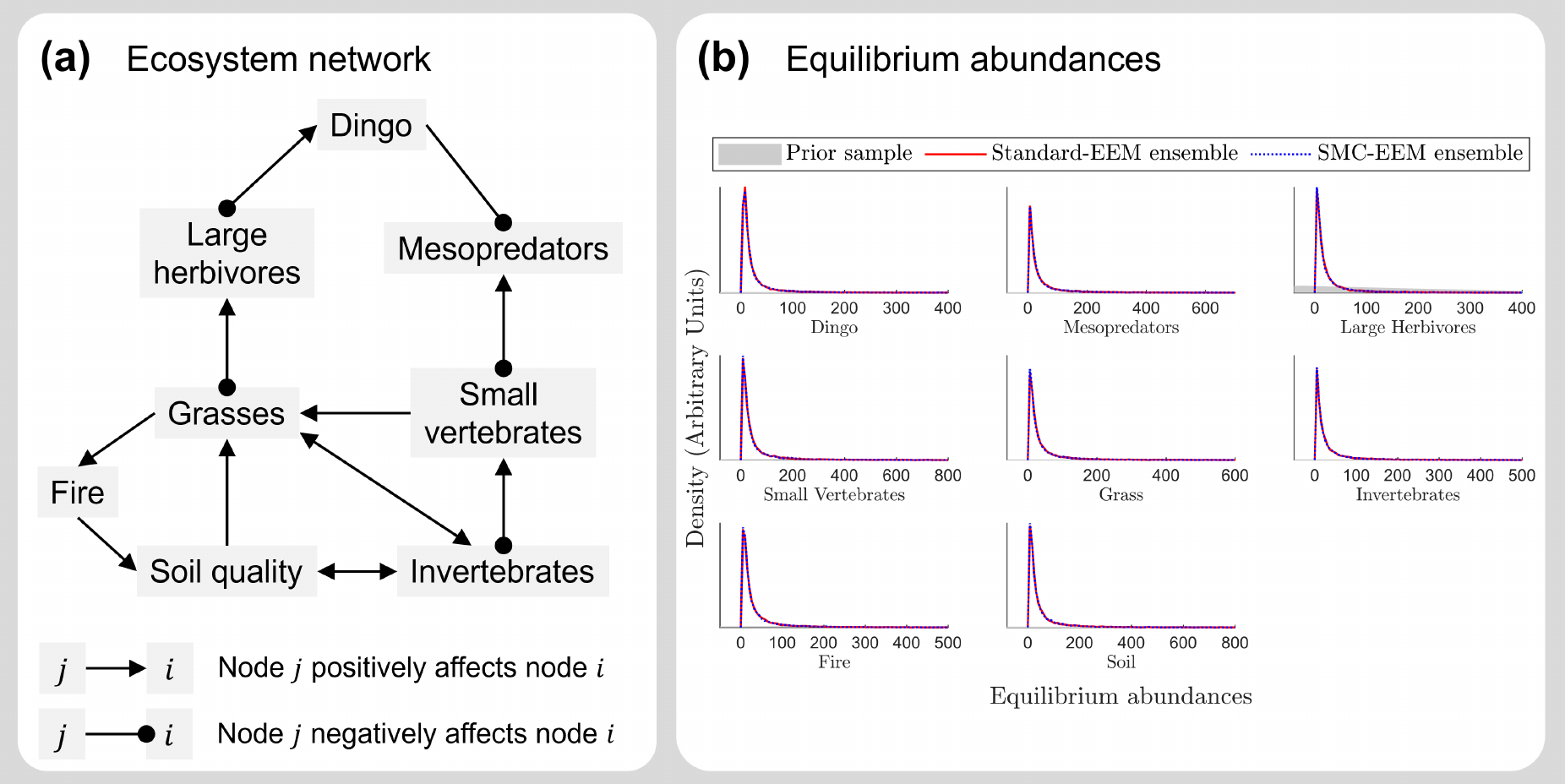}
\caption{{\bf Equilibrium abundances for the semiarid Australia network.}
Ensemble ecosystem modelling for an ecosystem network representing semiarid Australia parameterised using standard-EEM and SMC-EEM methods. \textbf{(a)} The semiarid Australian ecosystem network \cite{baker2017_EEM} consisting of eight nodes and 33 parameters when represented as a Lotka-Volterra system. {\edits \textbf{(b)} Distributions of equilibrium abundances from the prior distribution (grey), standard-EEM (red) and SMC-EEM (blue) ensembles of ecosystem models. Note that the range of equilibrium populations for the prior distribution is very diffuse (and hence barely visible in these plots) compared to the ensemble-predicted distribution abundances. Here the blue and red densities match almost exactly, demonstrating that the outputs of the standard-EEM and SMC-EEM methods are consistent. }}
\label{fig:dingo_square}
\end{figure}

For this network (Fig~\ref{fig:dingo_square}a), we found that the SMC-EEM method produced consistent {\edits estimated distributions of equilibrium abundances} to the standard-EEM method (Fig~\ref{fig:dingo_square}b). We also observe similar estimated parameters (Fig \ref{Sfig: semiarid marginals}), stiff eigenparameters (Fig \ref{Sfig: semiarid eigenparameters}), and {\edits time-series predictions (Fig \ref{Sfig: semiarid timeseries})} for the standard-EEM and SMC-EEM produced ensembles. 

\subsection*{Case study 2: Phillip Island network}
The standard-EEM method required 108 days to generate 100,000 ensemble members for the Phillip Island network; however, SMC-EEM completed this task in under 6 hours (Table \ref{tab:Phillip results}). {\edits (It should be noted that these computational exercises were performed in parallel on 12 cores.)} The SMC-EEM method produced the ensemble in 0.22\% of the time required by standard-EEM because it required 0.13\% of the simulations. This massive computational saving is consistent with the results presented in Fig \ref{fig:Computation time acceptance rate}, as the acceptance rate for the Phillip Island network was $1.7\times10^{-6}$. The SMC-EEM method is thus the only practical option, out of the two methods, for this 22-species network. 

\begin{table}[!ht]
\centering
\begin{tabular}{|l|c|c|}
\hline
 & \textbf{Standard-EEM} & \textbf{SMC-EEM}\\ \hline \hline
Computation time (sec) & 9.3$\times10^6$ & $2.1\times10^4$ \\ \hline
Simulations (number) & $5.8\times10^{10}$ & $7.8\times10^7$\\ \hline
\end{tabular}
\caption{{\bf Computational requirements for the Phillip Island network.} Computation time and the number of simulations required to generate an ensemble of 100,000 models using standard-EEM and SMC-EEM for the Phillip Island ecosystem network.}
\label{tab:Phillip results}
\end{table}

\begin{figure}
\centering
\includegraphics[width=0.8\textwidth]{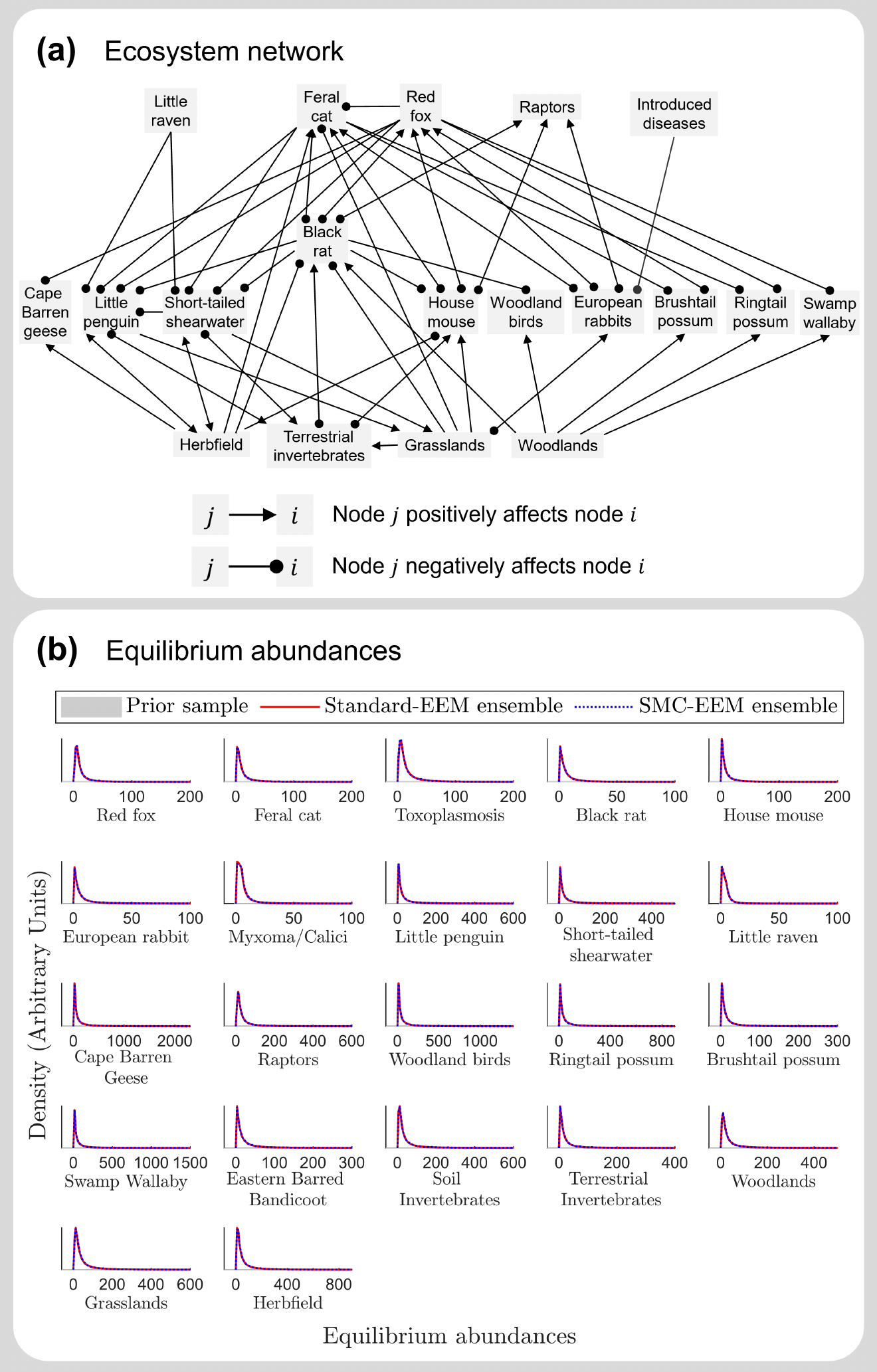}
\caption{{\bf Equilibrium abundances for the Phillip Island ecosystem network.} Ensemble ecosystem modelling for an ecosystem network representing Phillip Island parameterised using standard-EEM and SMC-EEM. \textbf{(a)} The Phillip Island ecosystem network \citep{Rendall_2021_EEMeg} consists of 22 nodes, with connectance $c=0.18$, and 110 parameters when represented as a Lotka-Volterra system. {\edits \textbf{(b)} Distributions of equilibrium abundances from the prior distribution (grey), standard-EEM (red) and SMC-EEM (blue) ensembles of ecosystem models. Note that the range of equilibrium populations for the prior distribution is very diffuse (and hence barely visible in these plots) compared to the ensemble-predicted distribution abundances. Here the blue and red densities match almost exactly, demonstrating that the outputs of the standard-EEM and SMC-EEM methods are consistent. }}
\label{fig:Phillip predictive distributions}
\end{figure}

Additionally, the outputs of SMC-EEM and standard-EEM are consistent. {\edits The distributions of equilibrium abundances computed for each parameterised ensembles are consistent (Fig~\ref{fig:Phillip predictive distributions}b) and both} methods produce comparable estimated marginal parameter distributions (Fig \ref{Sfig: Phillip marginals}), stiff eigenparameter distributions (Fig \ref{Sfig: Phillip eigenparameters}) {\edits and population forecasts (Fig \ref{Sfig: Phillip timeseries}),} indicating that the information gained about the parameters is consistent between methods.

\subsection*{Case study 3: Great Barrier Reef network}
Parameterising the Great Barrier Reef network \cite{Rogers_2015_reefNetwork} for 100,000 ensemble members took $21$ hours for the SMC-EEM method (Table \ref{tab:GBR results}) and could not be practically computed using the standard-EEM method. Based on a preliminary analysis of 20 ensemble members, it took approximately 40 hours to generate a single ensemble member using standard-EEM with an acceptance rate of $\mathcal{O} (10^{-9})$, hence an ensemble of this size would take years to produce (estimated 450 years). The SMC-EEM method is thus the only practical option, out of the two methods, for this 16-species network.
 
\begin{table}[!ht]
\centering
\caption{
{\bf Computational requirements for the Great Barrier Reef network.}}
\begin{tabular}{|l|c|c|}
\hline
 & \textbf{Standard-EEM} & \textbf{SMC-EEM}\\ \hline \hline
Computation time (sec) & $\mathcal{O} (10^{10})$ & $7.6\times10^4$ \\ \hline
Simulations (number) & $\mathcal{O} (10^{13})$ & $1.5\times10^8$\\ \hline
\end{tabular}
\begin{flushleft} Computation time and the number of simulations required to generate an ensemble of 100,000 models using SMC-EEM for the Great Barrier Reef ecosystem network. The standard-EEM method could not be used to generate an ensemble of this size within a practical time-frame, so the results in Table \ref{tab:GBR results} are estimated using {\edits an ensemble of 20 parameter sets}. 
\end{flushleft}
\label{tab:GBR results}
\end{table}

\begin{figure}[H]
\centering
\includegraphics[width=0.7\textwidth]{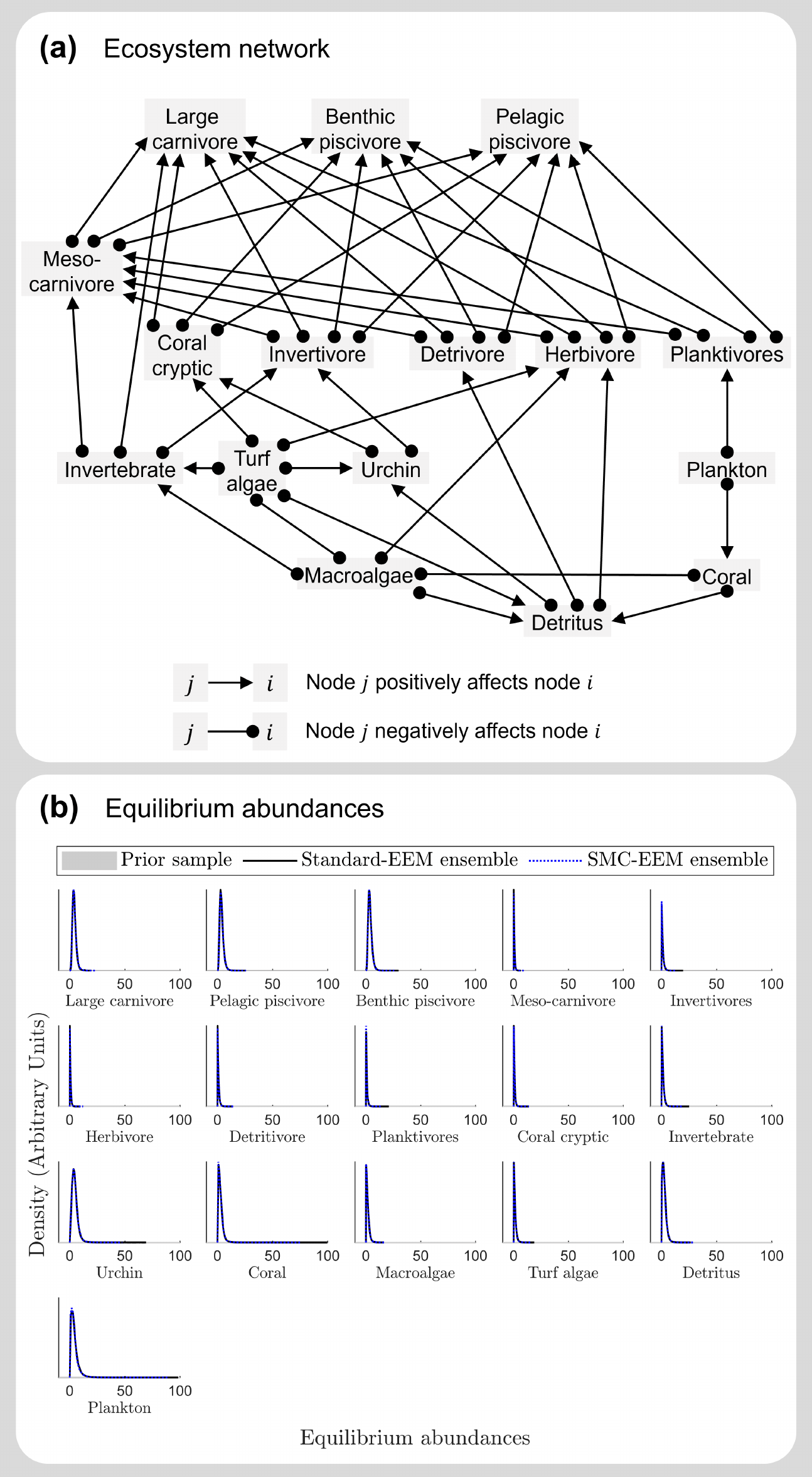}
\caption{{\bf Equilibrium abundances for the Great Barrier Reef network.} Ensemble ecosystem modelling for an ecosystem network representing the Great Barrier Reef parameterised using standard-EEM and SMC-EEM. \textbf{(a)} The Great Barrier Reef ecosystem network \citep{Rogers_2015_reefNetwork} consists of 16 nodes, with connectance $c=0.4$, and 118 parameters when represented as a Lotka-Volterra system. {\edits \textbf{(b)} Distributions of equilibrium abundances from the prior distribution (grey), and two independent SMC-EEM ensembles (light blue and dark blue) of ecosystem models. Note that the range of equilibrium populations for the prior distribution is very diffuse (and hence barely visible in these plots) compared to the ensemble-predicted distribution abundances. Here the independent SMC-EEM {\edits ensembles} are consistent, demonstrating reproducibility.}}
\label{fig:GBR predictive distributions}
\end{figure}

Since we cannot produce a standard-EEM {\edits ensemble}, instead we compared the outputs of two independently obtained SMC-EEM {\edits ensembles} to assess their reproducibility. This indicates if SMC-EEM can adequately sample the parameter space to produce a representative {\edits ensemble}. {\edits The two independent SMC-EEM ensembles of 100,000 yield consistent results when comparing the predicted equilibrium abundances (Fig~\ref{fig:GBR predictive distributions}b)}. Additionally, the {\edits ensembles} have comparable estimated marginal parameter distributions (Fig \ref{Sfig: GBR marginals}), stiff eigenparameter distributions (Fig \ref{Sfig: GBR eigenparameters}), {\edits and time-series forecasts (Fig \ref{Sfig: GBR timeseries}),} indicating that the information gained about the parameters is consistent across independent runs. Such a result is very encouraging given that, for this case study, we have yielded a representative approximation of 118-dimensional space with 100,000 {\edits parameter sets} each. 

{\edits Once an ensemble is obtained, a data-informed sensitivity analysis -- such as the analysis of model sloppiness -- can be used to identify the important parameter combinations for achieving feasibility and stability. For this Great Barrier Reef ecosystem network, each of the five most tightly constrained parameter combinations focuses on balancing the positive growth rates of basal species, or the self-regulation of top predators (see Fig \ref{Sfig: GBR 5 eigenparameters} and Fig \ref{Sfig: GBR 100 eigenparameters}). 

For example, using the information in Fig \ref{Sfig: GBR 5 eigenparameters}, the first eigenparameter can be expressed as $$\hat{\theta}_1 = \frac{r_{TA} \cdot \alpha_{MA,MA}^{0.1} \cdot \alpha_{MA,C}^{0.1} \cdot \alpha_{MA,D}^{0.1}} {r_{MA}^{0.6} \cdot \alpha_{TA,U}^{0.1} \cdot \alpha_{TA,MA}^{0.1} \cdot \alpha_{TA,D}^{0.1}} \approx \frac{r_{TA}}{\sqrt{r_{MA}}},$$
where $r_i$ is the positively constrained intrinsic growth rate for species $i$, $\alpha_{i,j}$ is the interaction parameter for the effect of species $j$ on species $i$, and the relevant species for this equation are represented as $TA$ for turf algae, $MA$ for macroalgae, $C$ for coral, $D$ for detritus and $U$ for urchins. This eigenparameter describes the balance between the proliferation of turf algae and the negative impacts on its abundance: mainly competition with macroalgae (including the proliferation rate of macroalgae), but also other lower trophic species including detritus, coral and urchins. Similar relationships can be seen for the five most influential parameter combinations (Fig \ref{Sfig: GBR 5 eigenparameters}). 

This could indicate that given growth rate parameters are constrained to be only positive, and self-interactions between species are constrained to be only negative (self-regulating), the most important features for parameterising feasible and stable ecosystems are a high abundance of basal species and limited populations of top-predators. This well-observed result, while not a surprising insight, indicates how this analysis could be used to identify key drivers for developing feasible and stable ecosystems. }

\section*{Discussion}
In this work, we have presented and demonstrated a method that, for the first time, can rapidly generate ensemble ecosystem models for higher dimensional ecosystem networks. This new method, which we call the SMC-EEM method, can generate consistent ensembles to the current gold-standard method -- standard-EEM -- whilst being orders of magnitude faster for large and densely connected networks. On a Phillip Island case study \cite{Rendall_2021_EEMeg} SMC-EEM reduced the computation time from 108 days to 6 hours, with indistinguishable time-series predictions, estimated distributions of model parameters and model parameter combinations. For a Great Barrier Reef network, we showed that standard-EEM was not capable of producing a large ensemble, such that SMC-EEM was the only practical option. This new method permits large and complex ecosystems -- as observed in nature -- to be practically simulated and analysed. 

\subsection*{The best ecosystem generation method depends on the properties of the ecosystem network}

Both the standard-EEM method and our introduced SMC-EEM method have advantages and disadvantages, depending on the ecosystem being modelled. SMC-EEM is expected to be more computationally efficient for ecosystems comprised of 7 or more species (result obtained for a connectance probability $c=0.5$ as in Fig~\ref{fig:Computation time comparison}; see Fig \ref{Sfig: connectances} for results with other values of $c$), or if less than 1 in 200 parameter values are feasible and stable when sampled using standard-EEM (acceptance rate of 0.005; Fig~\ref{fig:Computation time acceptance rate}). While the acceptance rate of an ecosystem network, which encapsulates both the number of species and connectance, is a better predictor of computation time than the number of species in the system (see Figs~\ref{fig:Computation time comparison} and \ref{fig:Computation time acceptance rate}), the number of species is a much more intuitive measure and does not require prior calculations to estimate, unlike the acceptance rate. 

When considering the eight-species semiarid Australian ecosystem network (with $c=0.39$), based on the number of species it would be unclear beforehand whether SMC-EEM or standard-EEM would be faster (Fig~\ref{fig:Computation time comparison}). However, by estimating the acceptance rate as $0.11$ (roughly 1 in 9 parameter sets tested were feasible and stable), Fig~\ref{fig:Computation time acceptance rate} clearly shows standard-EEM is expected to outperform SMC-EEM for this network. Practically, both standard-EEM and SMC-EEM are acceptable choices for this case study as they both generated the model ensemble within a minute; however, we must acknowledge that standard-EEM is {\edits a simpler process (making it more straightforward to implement in computer code)} and generated the ensemble faster (Table \ref{tab:semiarid results}).

In contrast, for the 22-species Phillip Island case study (with $c=0.18$) and an acceptance rate of $1.7\times10^{-6}$  (roughly 1 in 600,000 parameter sets tested were feasible and stable), it is clear from both Figs~\ref{fig:Computation time comparison} and \ref{fig:Computation time acceptance rate} that SMC-EEM will be significantly faster. When applying standard-EEM to this system, we found it would take 108 days to generate the ensemble (Table \ref{tab:Phillip results}), making SMC-EEM the only practical option of the two methods. 

Lastly, the 16-species Great Barrier Reef network (with $c=0.4$) and an acceptance rate of $\mathcal{O} (10^{-9})$ (roughly 1 in a billion parameter sets tested were feasible and stable) is expected to be orders of magnitude faster according to the trends shown in Figs~\ref{fig:Computation time comparison} and \ref{fig:Computation time acceptance rate}, and the observed computation times (Table \ref{tab:GBR results}) were within the credible ranges indicated by these trends. Here we note that the acceptance rate for this network is considerably smaller than for the Phillip Island network, and this could be due to being more densely connected, or the structure of the network itself \citep{Johnson_2014,Barbier_2019_pyramids}.

\subsection*{Comparing the ensembles generated by the two methods}

In this work, we used the estimated parameter distributions and time-series predictions to compare {\edits ensembles} produced using the two methods. Additionally, the distributions of the stiff eigenparameters, obtained using an analysis of model sloppiness, provided an additional diagnostic comparing the similarity of the {\edits ensembles}. The analysis of model sloppiness can indicate how similar the {\edits ensembles} are, whilst accounting for parameter interdependencies \citep{Monsalve_2022,botha_2022} -- a perspective not easily observed via the estimated marginal distributions, {\edits quantities of interest,} or via time-series predictions. We, therefore, encourage the comparison of Bayesian inference method-generated {\edits ensembles} via comparison of eigenparameter distributions alongside a comparison of marginal parameter distributions, as this provides a more comprehensive comparison. For the {\edits ensembles} tested in this work, the eigenparameter distributions did not indicate any substantive differences (Figs~\ref{fig:RMT example outputs},~\ref{Sfig: semiarid eigenparameters}~and~\ref{Sfig: GBR eigenparameters}).

To our best knowledge, the SMC-EEM method outputs match those produced by the standard-EEM method (Figs~\ref{fig:RMT example outputs}--\ref{fig:GBR predictive distributions}, and \ref{Sfig: semiarid marginals}--\ref{Sfig: GBR eigenparameters}). However, users should be cautious when selecting the ensemble size for SMC-EEM. {\edits While standard-EEM always randomly samples from the parameter space to propose new values, SMC-EEM proposes new values relative to current values in the ensemble (via the multivariate Gaussian proposal distribution centered on the current parameter value within the MCMC algorithm). Hence, if} there are not enough {\edits particles} to cover a high-dimensional parameter space, the SMC-EEM method may not sufficiently explore the parameter space, thereby creating an ensemble that is not representative and is different to the distribution of ensembles produced by standard-EEM. This difference in ensembles occurred when using only 10,000 ensemble members for both the Phillip Island and Great Barrier Reef case studies; however, the ensembles were found to be consistent for 100,000 ensemble members. 

For ecosystem networks that are not overly complex, it is possible to assess whether there are enough {\edits parameter sets} by comparing the results of SMC-EEM and standard-EEM. But for high-dimensional ecosystem networks, it will not be practical to compare outcomes since the latter will have impractically high computational costs (as for the Great Barrier Reef case study). We therefore recommend multiple independent runs of the SMC-EEM method and a visual assessment of whether the ensemble is reproducible (through the estimated parameter distributions, stiff eigenparameter distributions, and time-series predictions), especially if the ecosystem network is as large as the Great Barrier Reef network explored here (see Fig \ref{fig:GBR predictive distributions}). Hence, while the foundational analysis presented here demonstrates that the SMC-EEM method finally unlocks analysis of higher-dimensional networks, its accuracy will be limited primarily by the size of the ensemble.  

\subsection*{Implications for ecosystem network generation in nature}

{\edits While the main motivation behind SMC-EEM was to maximise the capabilities of the conservation tool, this parameterisation regime could also be of use for drawing theoretical insights.} There is substantial debate in the literature regarding which features of natural ecosystems make them more likely to be stable and feasible (e.g., \cite{emmerson_2004,Allesina_2012_stab,jacquet_2016}). Some literature suggests that larger and more connected networks are less likely to be feasible and stable \citep{May_1972,Dougoud_2018,Allesina_2012_stab} because there is a lower probability of randomly sampling parameter values to satisfy these two constraints. However, treating the probability of generating a feasible and stable system through random sampling as a proxy for the likelihood of these systems developing in nature creates a disparity: complex food webs are actually observed in nature, yet are perceived theoretically as highly unlikely. 

{\edits Interactions in ecosystems have been shaped by processes such as co-evolution, niche partitioning, and resource competition \cite{dormann_2017}, making it unlikely that interactions in ecological networks are random. Additionally,} the ``community assembly'' hypothesis \citep{weiher_2011} suggests that the development and persistence of large food webs may be the result of natural selection of species survival (from an even larger pool of initial species) whose interaction strengths possess particular statistical properties \citep{barbier_2021,Servan_2018}. {\edits These theories imply} that the probability of randomly sampling parameter values to satisfy feasibility and stability does not indicate the probability of the ecosystem existing in nature. 

Thus, instead of being limited by the conceptual argument that the inability to efficiently generate plausible ecosystems via random sampling suggests these ecosystems cannot exist in practice, a key implication of the community assembly hypothesis is that we can instead take advantage of the full suite of Bayesian approaches (as performed here) to identify an ensemble of parameters that can plausibly generate large ecosystems in a computationally efficient manner. The SMC-EEM method also has the potential (beyond specific case studies) to broadly explore the consequences of community assembly on the general properties of ecosystem networks that form in nature \cite{barbier_2021}. 

{\edits Now that we can quickly produce large ensembles of parameter values that match ecological theory, insights can be drawn from the results. This method could be used to compare the relative difficulties in obtaining models that meet different constraints; for example, is there a lower probability of obtaining feasible ecosystem models, or stable ecosystem models? Alternatively, practitioners could compare the estimated parameter values, or values of interest -- such as abundance correlations between species -- across ensembles parameterised using different ecological theories. 

In our implementation, we assumed parameters were independent in the prior distribution; however, SMC-EEM can accommodate other prior choices (e.g., prior parameter dependencies such as a trophic transfer efficiency constraint \cite{baker2017_EEM} or intraspecific density dependencies \cite{Reznick_2002} can be implemented using conditional distributions). However, assuming prior parameter independence does not prevent dependencies from being inferred when fit to the constraints. By analysing the covariance of the parameters once incorporating the constraints (using a method such as the analysis of model sloppiness), the parameter combinations that are important for feasibility and stability could be assessed, as we have shown in our analysis. 

When we applied this analysis to the Great Barrier Reef case study, it suggested that high populations of basal species and low populations of top predators were the most important factors for achieving the constraints. While this result is unsurprising, it is also somewhat uninsightful. This is likely due to the relatively uninformed prior distributions used in the analysis (following those of Baker et al., \cite{baker2017_EEM}) that forced intrinsic growth rate parameters to be positive and had equal magnitude across all species. Growth rate prior distributions with negative values, or other prior distributions, could easily be used within SMC-EEM instead. However, any effect of these prior distributions on the ensemble would in turn affect this analysis, such that we recommend testing various prior specifications to assess its impact. }

\subsection*{Computational efficiency unlocks new opportunities for improving ecosystem model realism} 

In the present analysis, we considered ecosystem networks generated by generalised Lotka-Volterra equations -- as this is the mathematical model that EEM has been thus far applied to \citep{baker2017_EEM} -- however, alternative models have been proposed to offer more complex representations of ecosystem interactions in nature, such as {\edits different functional responses} \citep{holling_1959}, or more recently, higher-order (i.e. beyond pairwise) interactions \citep{Gibbs_2022}. The generalised Lotka-Volterra model is computationally convenient for EEM because the equilibrium feasibility and stability conditions are readily computable via algebraic formulae (Equations \eqref{eq: equilibrium matrix} and \eqref{eq:jacobian}). A different choice of model or constraints could be much more computationally expensive to simulate and include many more parameters for calibration -- e.g., models with predator learning or prey saturation \citep{holling_1959}, or constraints on ecosystem dynamics outside of the system equilibrium \citep{Hastings_2018_transient}. The statistical efficiency of the SMC-ABC-based approach underlying our SMC-EEM method therefore offers a significant advantage over standard-EEM if other (potentially more realistic) model types and constraints are used. We surmise that the computational gains shown in the present work are expected to extend beyond the generalised Lotka-Volterra models, and feasibility and stability constraints considered here.  

Within our SMC-EEM method, the choice of discrepancy function drastically reduced the computation time in comparison to the standard-EEM method for larger networks (Fig~\ref{fig:Computation time acceptance rate}). We used a simple discrepancy function to indicate a measure of how infeasible and unstable an ecosystem parameterisation is (Equation \eqref{eq: discrepancy}); however, there may be better choices for the discrepancy function which further improve the efficiency of the method -- such as replacement of the sums and absolute values in Equation \eqref{eq: discrepancy} with other distance measures like the Euclidean norm, or weighting the infeasibility and instability sums differently. We leave these investigations for future work, especially as the results regarding the ``best'' discrepancy function may be highly model and constraint-specific. 

When additional constraints are imposed on the ensemble -- which further reduces the acceptance rate -- maintaining computational efficiency carries even greater importance than seen here. Case studies in the literature have considered constraints in addition to feasibility and stability, including feasibility and stability for subsets of the ecosystem \citep{baker2017_EEM,Peterson_2021_DirkHartog}, randomly assigned species interactions \citep{Rendall_2021_EEMeg,Peterson_2021} and additional constraints on combinations of parameters (e.g.\ trophic energy transfer constraints) \citep{baker2017_EEM,Peterson_2021_DirkHartog}. {\edits While the inclusion of such additional constraints in the SMC-EEM method is possible, it can require more careful algorithmic programming than the standard-EEM method.}

{\edits Additional data on population estimates, where available, should be used to inform the model parameters further. Since the constraints we used to parameterise SMC-EEM are not directly observable, we can consider the resulting ensemble as a constraint-informed prior distribution \citep{Wesner_2021} which can then be updated to incorporate any available time-series data in a subsequent Bayesian analysis. Furthermore, it would be interesting to analyse the effects on population forecasts of the constraint-informed prior compared to the relatively uninformed prior. Alternatively, the constraints within the discrepancy function could be redefined where additional information about species abundance estimates is available (see e.g., Neutel et al \cite{neutel_2007_data_snapshot}). Parameter sets with equilibrium abundances near the estimates could be given a lower discrepancy according to a Gaussian distribution, or equilibrium abundance limits could be defined -- as in the feasibility constraint (see Equation \eqref{eq: discrepancy feasibility}) -- to avoid unreasonable population sizes. Though connecting these data with feasibility and stability constraints, we hope that ensemble ecosystem modelling can be more accurate for conservation decision-making. }

\section*{Conclusion}
Through SMC-EEM we have unlocked ensemble ecosystem modelling for large and complex networks. Increasing the computational efficiency means that users only need to wait hours, rather than months, to analyse the risks and potential consequences of conservation actions in remote and understudied ecosystems with limited data. Through drastically improved computational efficiency, SMC-EEM brings new opportunities to explore more realistic ecosystem models and constraints to study the large and complex ecosystem networks that exist in nature. 

\section*{Supporting information}

\paragraph*{S1 Code.}
\label{Code}
{\bf Software for using the methods and reproducing the results presented in the manuscript.} The code used for this analysis was implemented in MATLAB (R2022b) and is freely available for download on Figshare at \url{https://doi.org/10.6084/m9.figshare.23707119.v2}. The code (495MB) 
associated with this manuscript contains an `EEM\_Methods' folder for using the methods and a separate `Results\_Replication' folder for reproducing the results and figures. 

\renewcommand\thefigure{S\arabic{figure}} 
\setcounter{figure}{0}  

\begin{figure}[H]
    \centering
    \includegraphics[width=\textwidth]{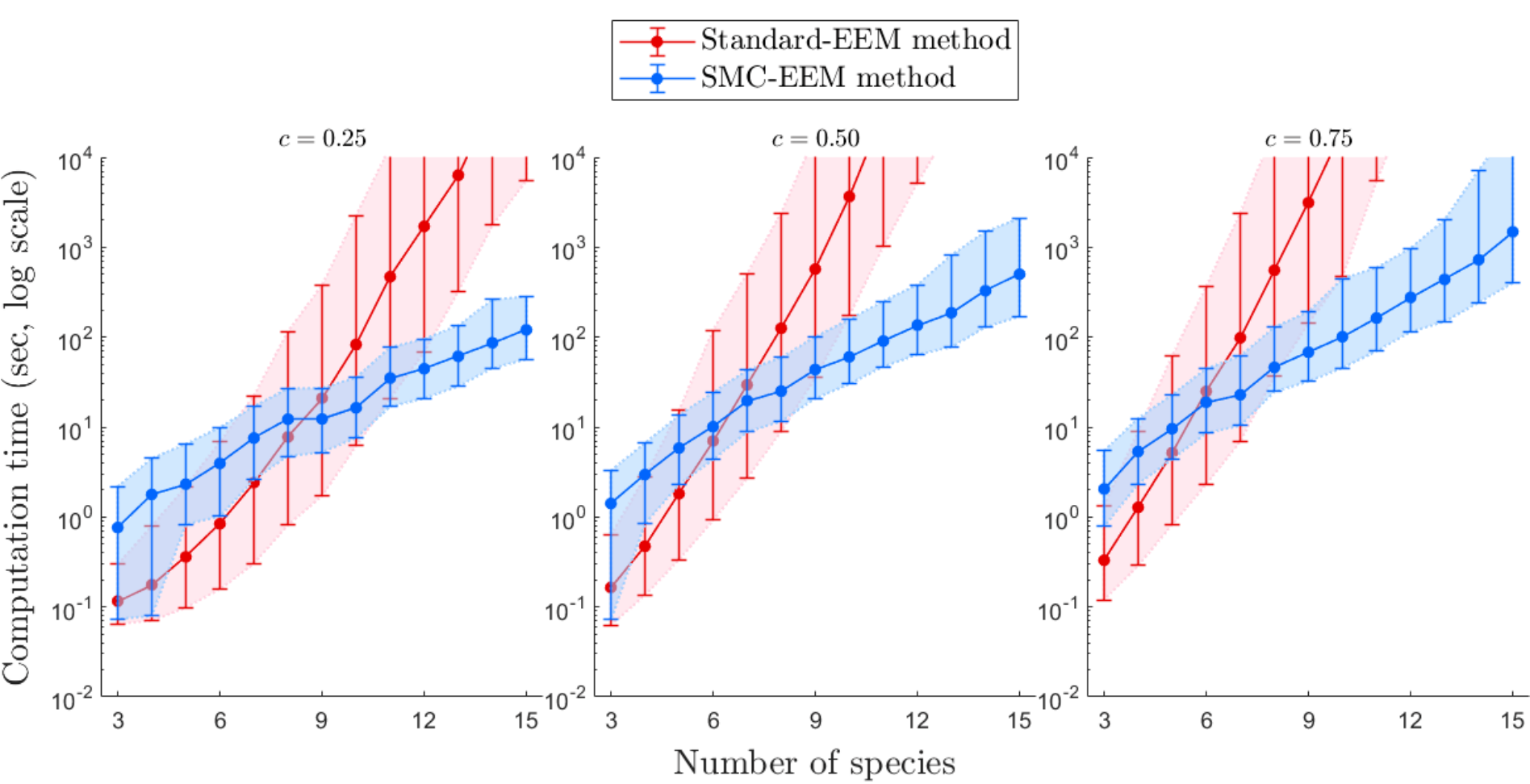}
    \caption{{\bf Computation time required to generate an ensemble for various network connectances.} The computation time needed to generate an ensemble of 1000 feasible and stable ecosystem models using a connectance probability of $c=0.25$ (left), $c=0.5$ (middle) and $c=0.75$ (right), for both the standard-EEM and SMC-EEM methods. This figure shows the medians (dots) and 7.5--92.5\% quantiles (error bars) of computation times for producing the results. {\edits Note, the computation time for any one ecosystem network was capped at $10^4$ seconds due to the computational burden of the simulation study.} More densely connected ecosystems (higher value of $c$) increase the computation time of both methods and decrease the network size at which the SMC-EEM method becomes more computationally efficient than the standard-EEM method. }
    \label{Sfig: connectances}
\end{figure}

\begin{figure}[H]
    \centering
    \includegraphics[width=0.8\textwidth]{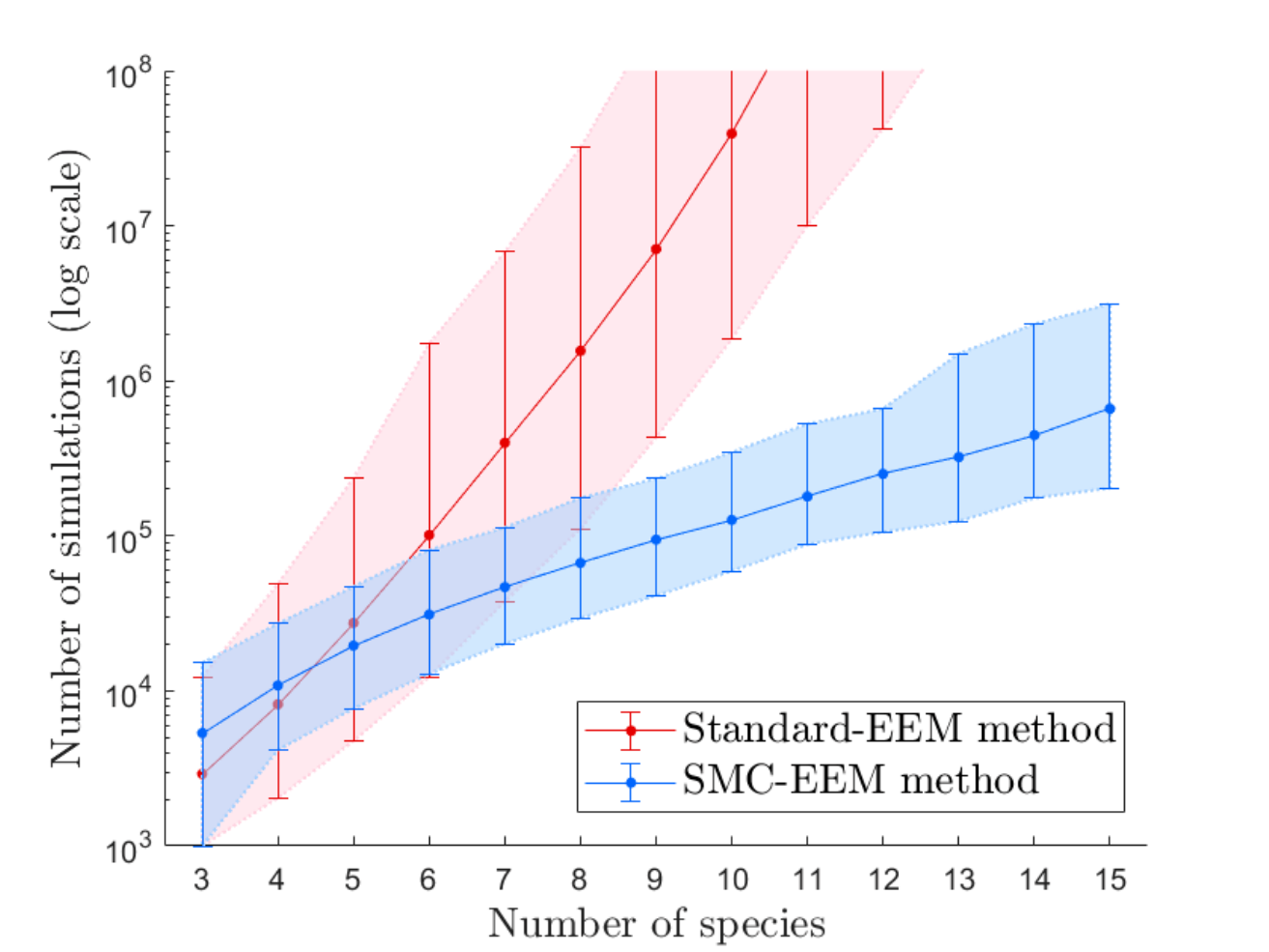}
    \caption{{\bf The number of simulations required to generate an ensemble for various network sizes.} The number of parameter sets trialled to generate an ensemble of 1000 feasible and stable ecosystem models using both the standard-EEM and SMC-EEM parameterisation methods. This figure shows the medians (dots) and 7.5--92.5\% quantiles (error bars) of simulation numbers for the models parameterised in Fig~\ref{fig:Computation time comparison} of the manuscript. {\edits Note, the computation time for any one ecosystem network was capped at $10^4$ seconds due to the computational burden of the simulation study.}}
    \label{Sfig: number of simulations}
\end{figure}

\begin{figure}[H]
    \centering
    \includegraphics[width=\textwidth]{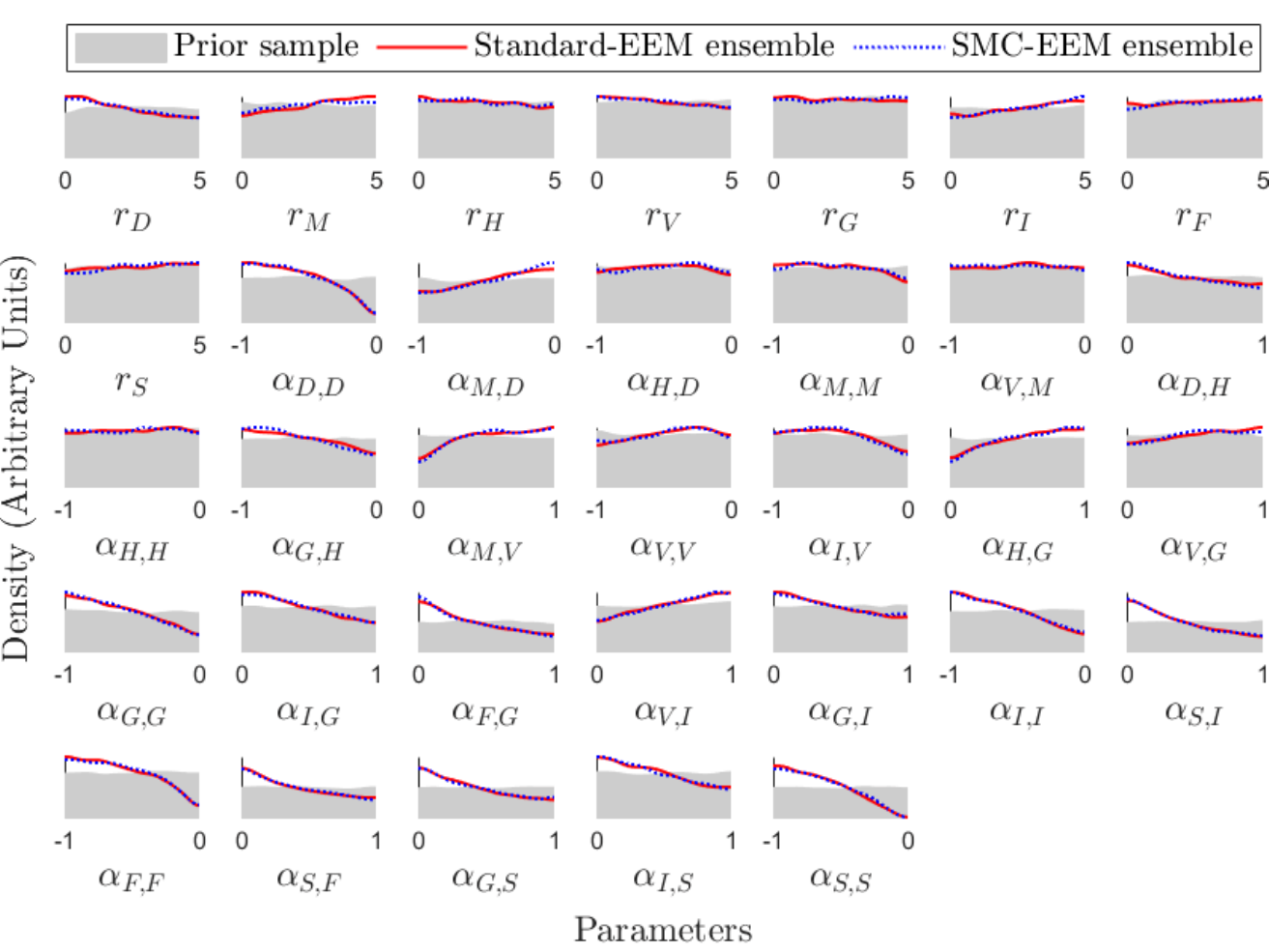}
    \caption{{\bf Parameter distributions for the semiarid Australia ecosystem network comparing standard-EEM to SMC-EEM.} Marginal parameter distributions estimated using both the standard-EEM method (red) and the SMC-EEM method (blue). Species labels represent dingoes (D), mesopredators (M), large herbivores (H), small vertebrates (V), grasses (G), invertebrates (I), fires (F) and soil quality (S). Notice that the blue and red densities match almost exactly, demonstrating that the outputs of the standard-EEM and SMC-EEM methods are consistent.}
    \label{Sfig: semiarid marginals}
\end{figure}

\begin{figure}
    \centering
    \includegraphics[width=0.9\textwidth]{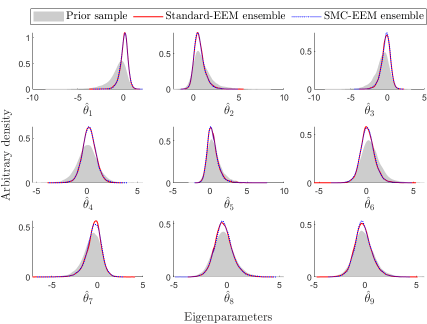}
    \caption{{\bf Eigenparameter distributions for the semiarid Australia ecosystem network comparing standard-EEM to SMC-EEM.} Marginal distributions of the nine stiffest eigenparameters estimated via the prior (grey), standard-EEM (red) and SMC-EEM (blue) ensembles. Notice that the blue and red densities match almost exactly, demonstrating that the outputs of the standard-EEM and SMC-EEM methods are consistent.}
    \label{Sfig: semiarid eigenparameters}
\end{figure}

\begin{figure}
    \centering
    \includegraphics[width=0.9\textwidth]{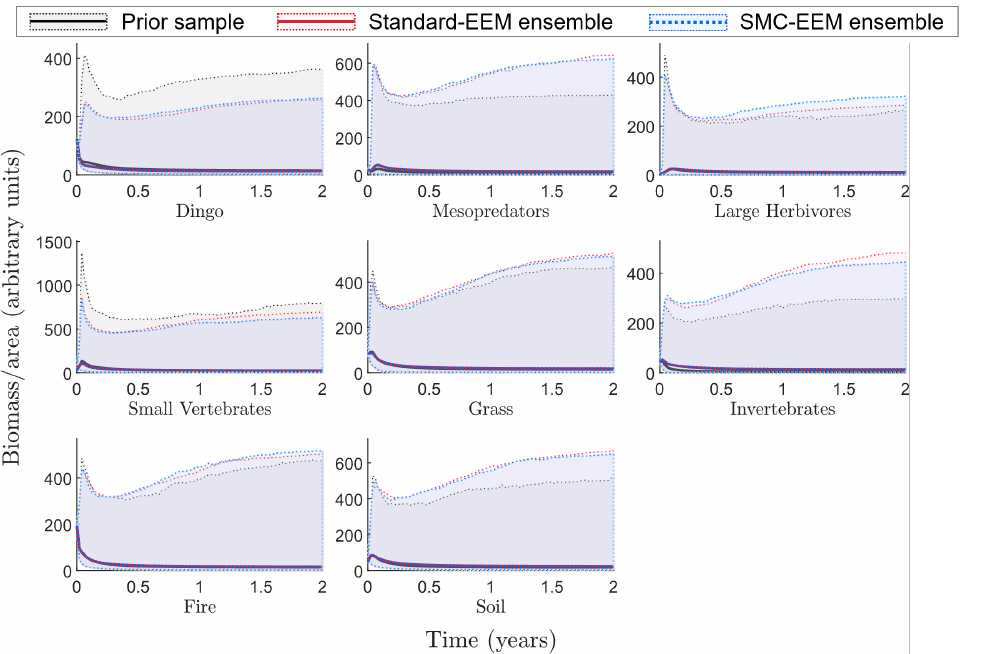}
    \caption{{\edits {\bf Time-series predictions for the semiarid Australia ecosystem network comparing the prior, standard-EEM, and SMC-EEM.} Time-series forecasts for the prior (grey), standard-EEM (red) and SMC-EEM (blue) ensembles simulated from a random initial condition. Depicted are the median (think lines) and 95\% credible intervals (thin dotted lines) for each {\edits ensemble}. Notice that the blue and red predictions are similar, demonstrating that the outputs of the standard-EEM and SMC-EEM methods are consistent.}}
    \label{Sfig: semiarid timeseries}
\end{figure}

\begin{figure}[H]
    \centering
    \includegraphics[width=\textwidth]{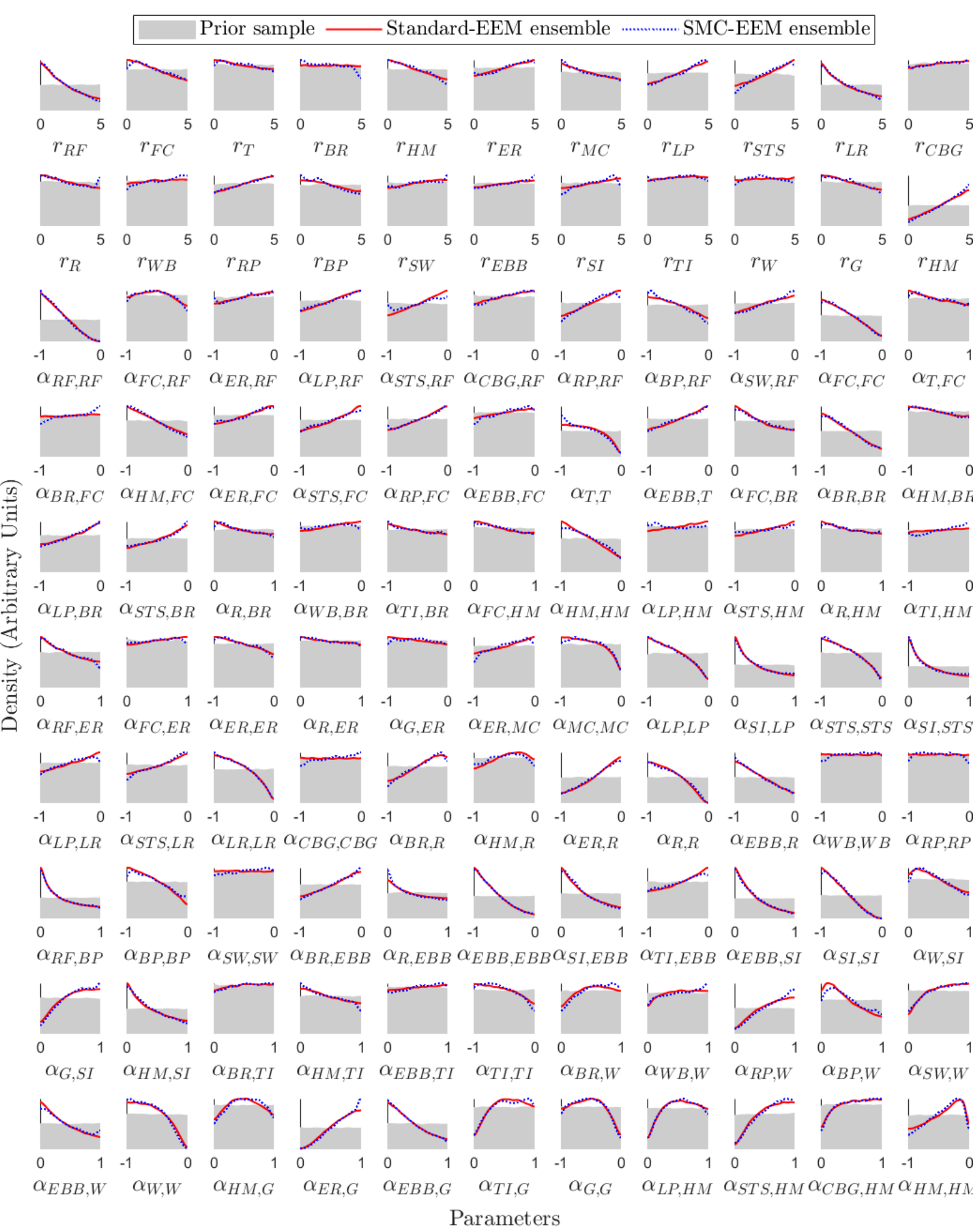}
    \caption{{\bf Parameter distributions for the Phillip Island ecosystem network comparing standard-EEM to SMC-EEM.} The estimated marginal distributions for each parameter within the ecosystem model for the Phillip Island network were generated via the standard-EEM method (red) and the SMC-EEM method (blue). Species labels represent parameters for the red fox (RF), feral cat (FC), toxoplasmosis (T), black rat (BR), house mouse (HM), European rabbit (ER), myxoma and calici (MC), little penguin (LP), short-tailed shearwater (STS), little raven (LR), Cape Barren geese (CBG), raptors (R), woodland birds (WB), ringtail possum (RP), brushtail possum (BP), swamp wallaby (SW), eastern barred bandicoot (EBB), soil invertebrates (SI), terrestrial invertebrates (TI), woodlands (W), grasslands (G), and herbfield (H).}
    \label{Sfig: Phillip marginals}
\end{figure}
\newpage

\begin{figure}[H]
    \centering
    \includegraphics[width=0.9\textwidth]{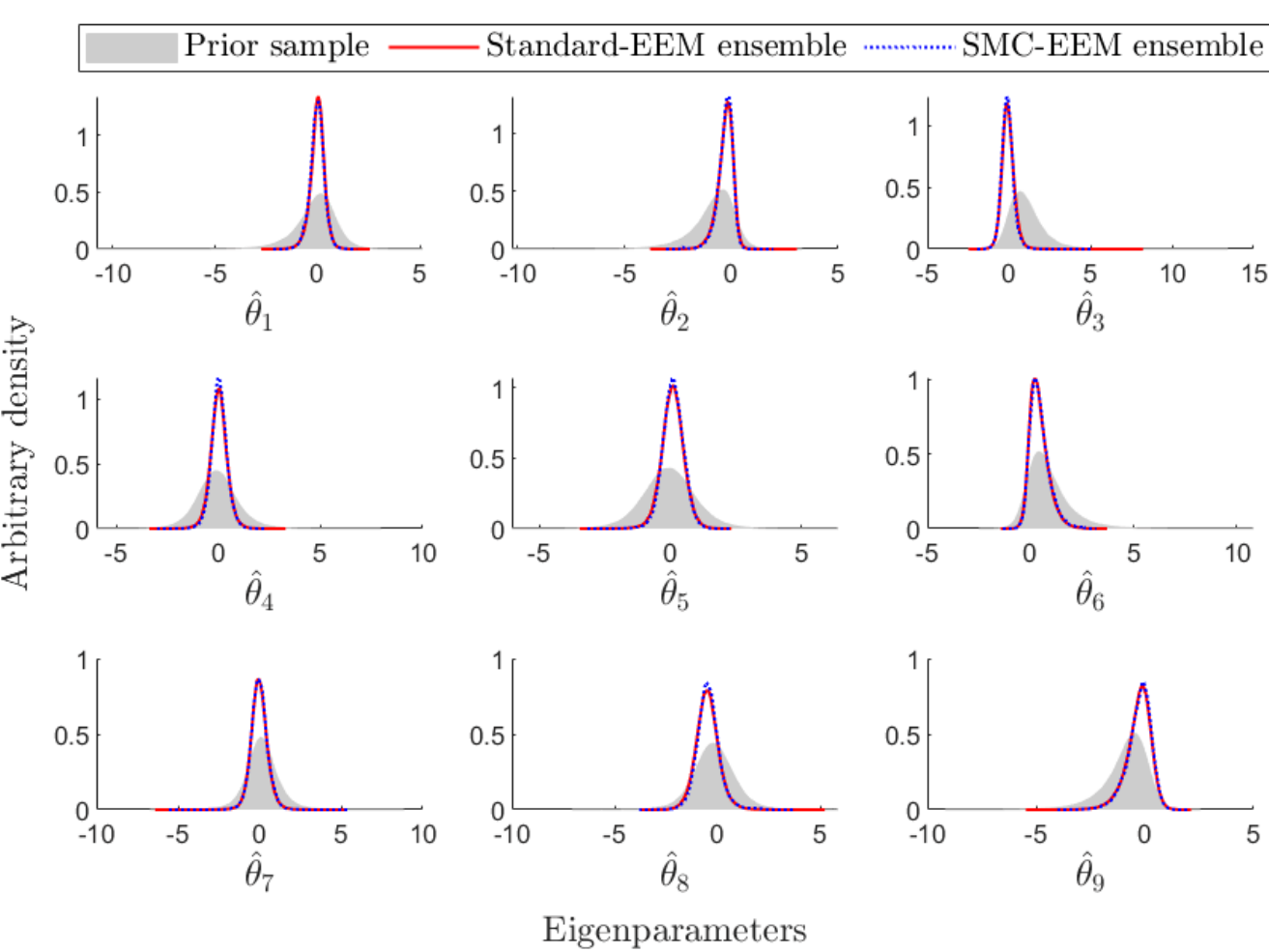}
    \caption{{\bf Eigenparameter distributions for the Phillip Island ecosystem network comparing standard-EEM to SMC-EEM.} Distributions of the nine most constrained parameter combinations (stiffest eigenparameters) determined by an analysis of model sloppiness of the standard-EEM ensemble. Here we compare the values of the eigenparameters for the prior (grey), standard-EEM (red) and SMC-EEM (blue) ensemble.}
    \label{Sfig: Phillip eigenparameters}
\end{figure}

\begin{figure}
    \centering
    \includegraphics[width=0.9\textwidth]{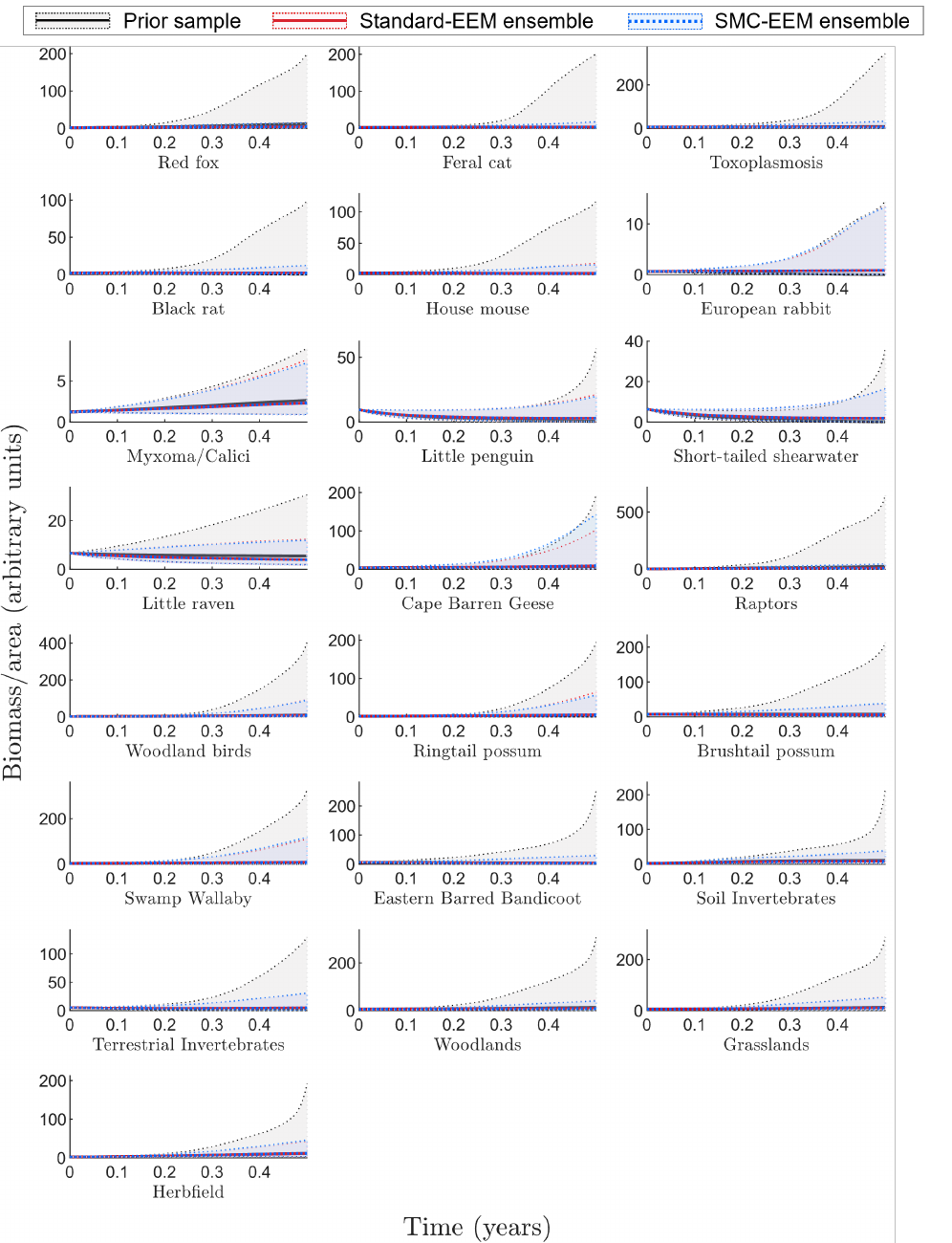}
    \caption{{\edits {\bf Time-series predictions for the Phillip Island ecosystem network comparing the prior, standard-EEM, and SMC-EEM.} Time-series forecasts for the prior (grey), standard-EEM (red) and SMC-EEM (blue) ensembles simulated from a random initial condition. Depicted are the median (think lines) and 95\% credible intervals (thin dotted lines) for each {\edits ensemble}. Notice that the blue and red predictions are similar, demonstrating that the outputs of the standard-EEM and SMC-EEM methods are consistent.}}
    \label{Sfig: Phillip timeseries}
\end{figure}

\begin{figure}[H]
    \centering
    \includegraphics[width=\textwidth]{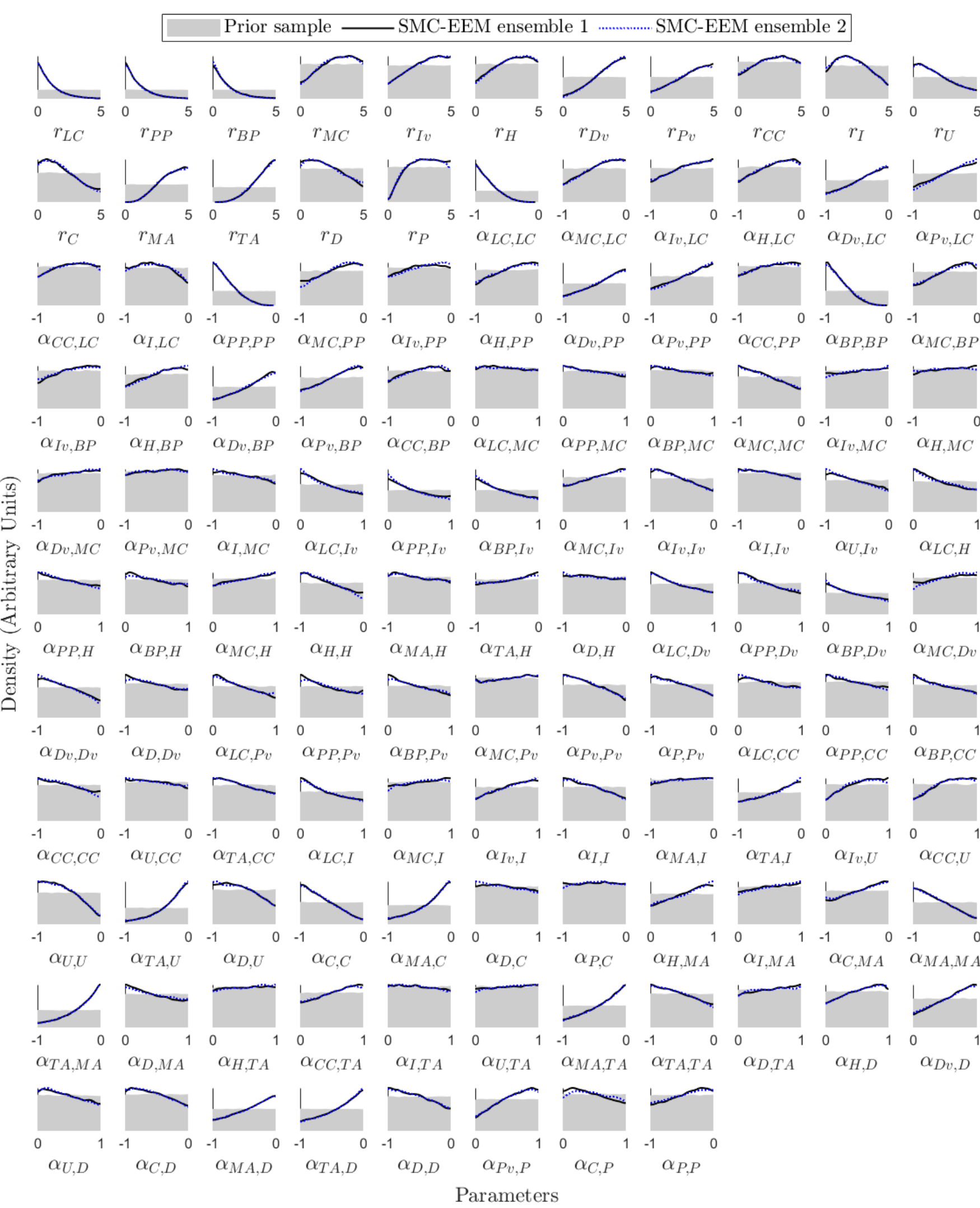}
    \caption{{\bf Parameter distributions for the Great Barrier Reef ecosystem network comparing two independent SMC-EEM ensembles.} The estimated marginal distributions for each parameter within the ecosystem model for the Great Barrier Reef network were generated via two independent runs of the SMC-EEM algorithm (black and blue). Species labels represent parameters for large carnivores (LC), pelagic piscivores (PP), benthic piscivores (BP), meso-carnivores (MC), invertivores (Iv), herbivore (H), detritivores (Dv), planktivores (Pv), coral cryptics (CC), invertebrates (I), urchins (U), corals (C), macroalgae (MA), turf algae (TA), detritus (D), and plankton (P). }
    \label{Sfig: GBR marginals}
\end{figure}

\begin{figure}[H]
    \centering
    \includegraphics[width=0.9\textwidth]{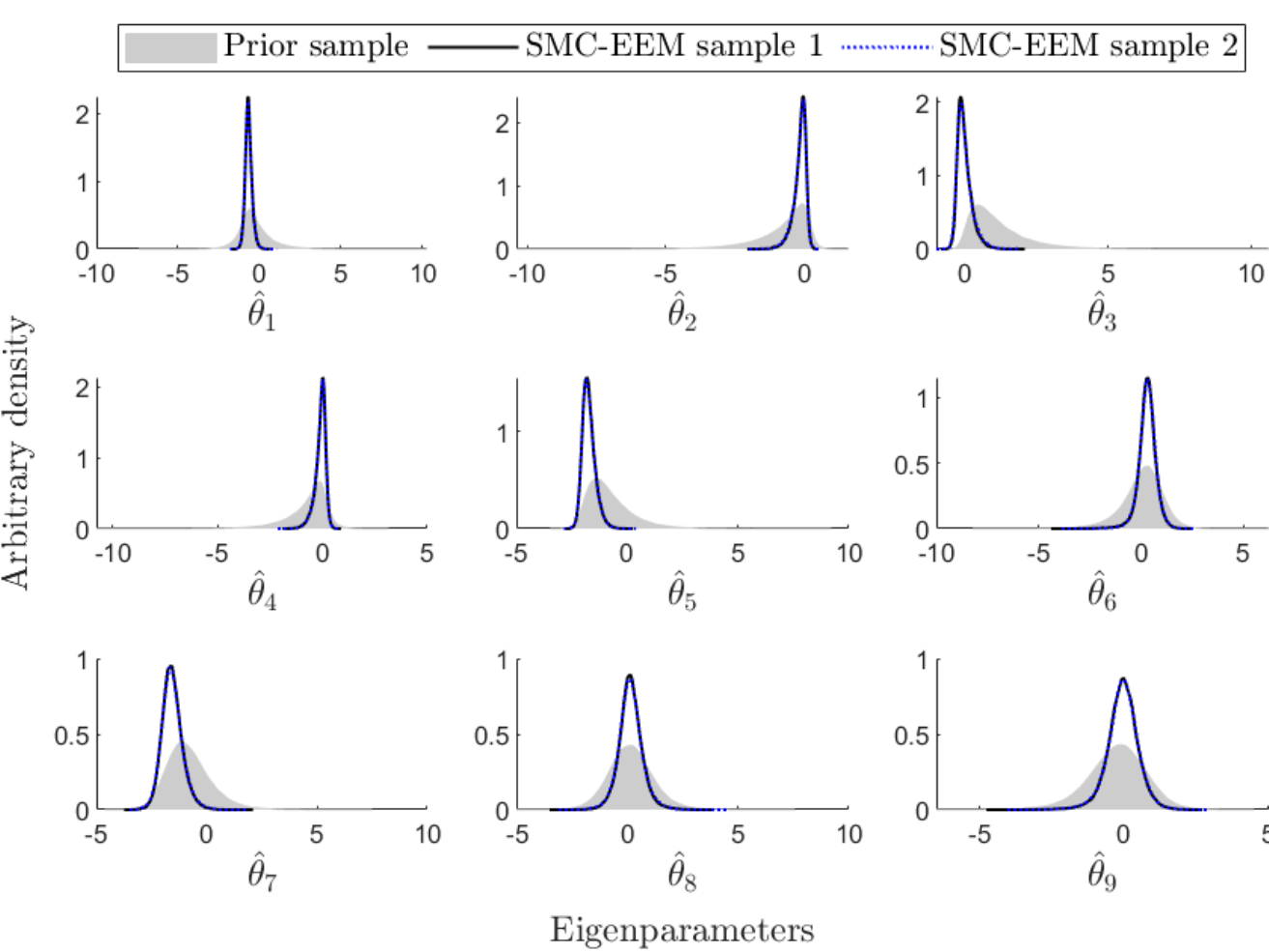}
    \caption{{\bf Eigenparameter distributions for the Great Barrier Reef ecosystem network comparing two independent SMC-EEM ensembles.} Distributions of the nine most constrained parameter combinations (stiffest eigenparameters) determined by an analysis of model sloppiness of a SMC-EEM ensemble. Here we compare the values of the eigenparameters for the prior distribution (grey), and two independent ensembles generated via the SMC-EEM algorithm (black and blue).}
\label{Sfig: GBR eigenparameters}
\end{figure}

\begin{figure}[H]
    \centering
    \includegraphics[width=0.9\textwidth]{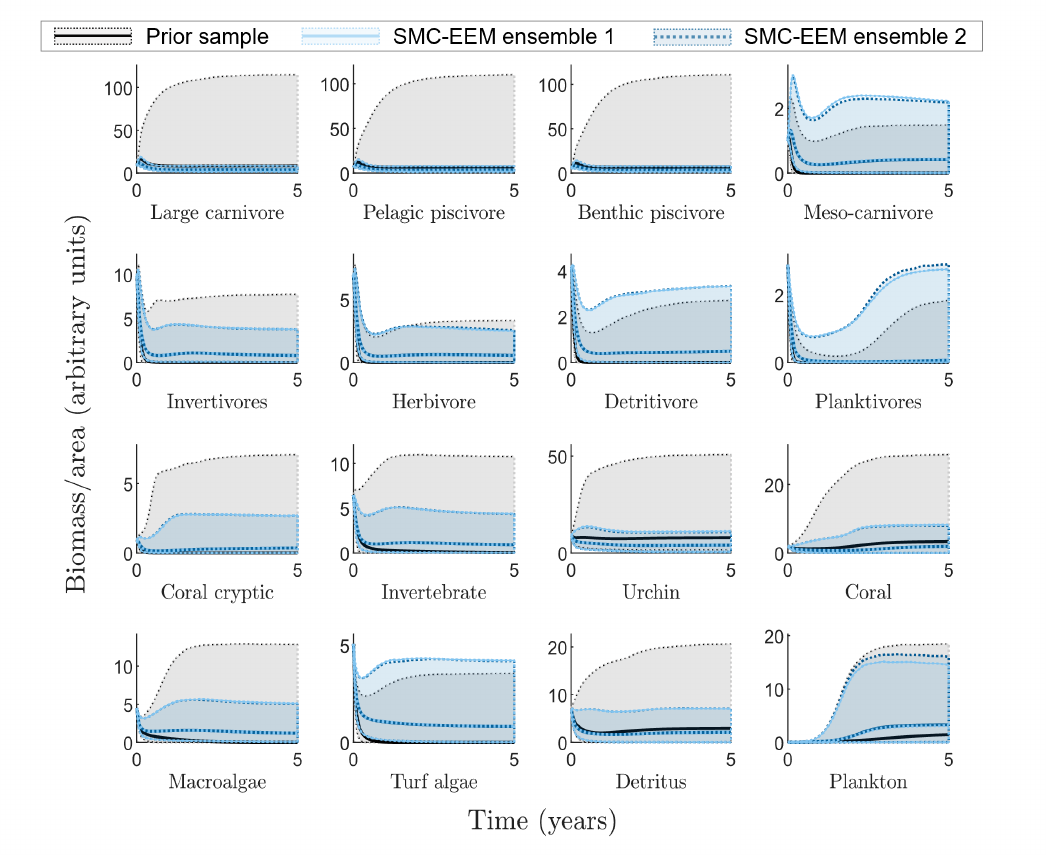}
    \caption{{\edits {\bf Time-series predictions for the Great Barrier Reef ecosystem network comparing the prior and two independently generated SMC-EEM ensembles.} Time-series forecasts for the prior (grey), and two independently generated SMC-EEM (light and dark blue) ensembles simulated from a random initial condition. Depicted are the median (think lines) and 95\% credible intervals (thin dotted lines) for each ensemble. Notice that the two blue predictions are similar, demonstrating that the SMC-EEM ensembles are consistent.}}
    \label{Sfig: GBR timeseries}
\end{figure}

\begin{figure}[H]
    \centering
    \includegraphics[width=\textwidth]{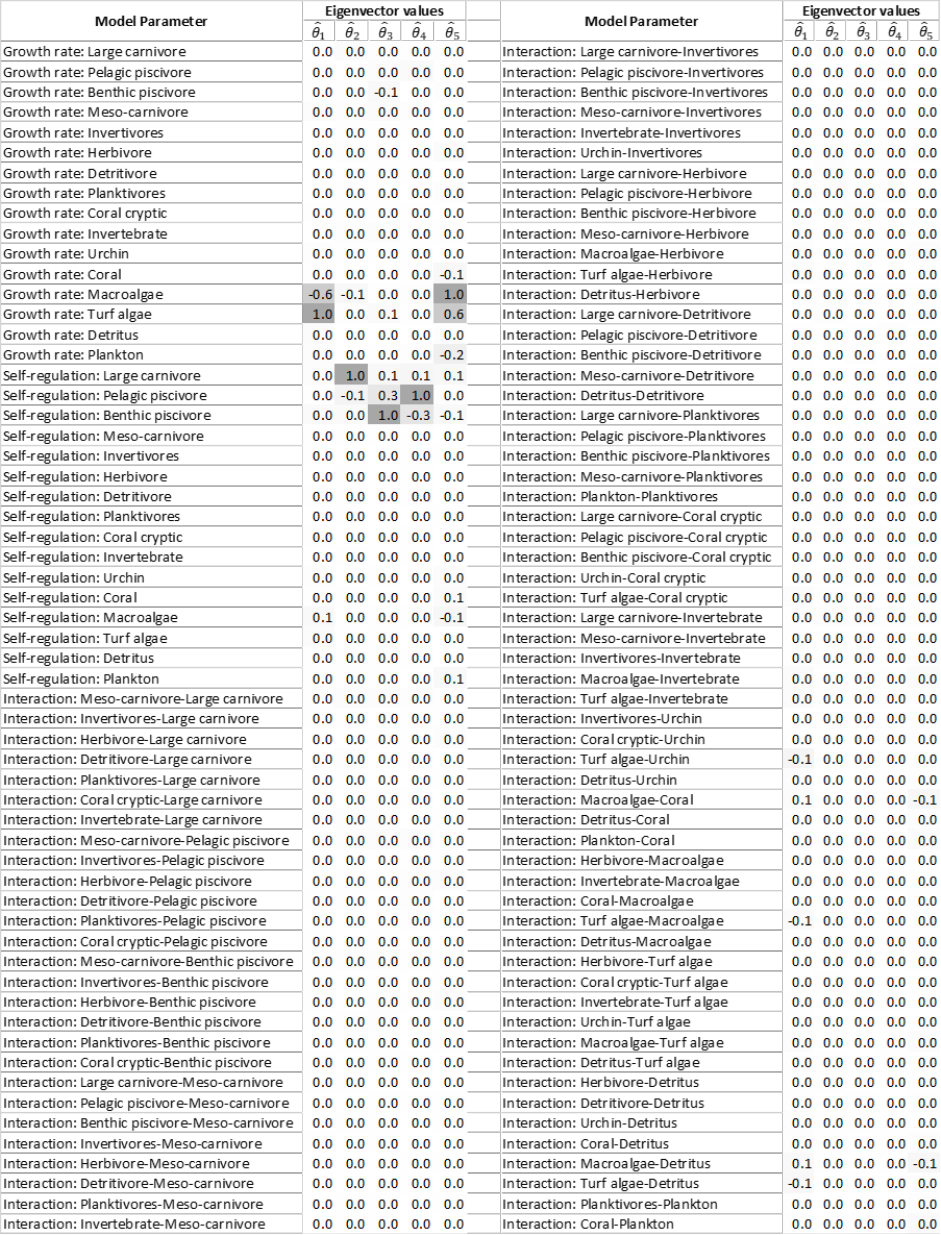}
    \caption{{\edits \textbf{Five most tightly constrained parameter combinations for the Great Barrier Reef ecosystem network. } The eigenvector values for the first five eigenparameters, rescaled to be between -1 and 1. These values are shaded such that the darker colours indicates a greater contribution of the parameter to the important parameter combinations. The columns of this table can be interpreted using Equation \eqref{Eq:Eigenparameter}. Notice, that the most important parameters are all growth rates for lower trophic species, and self-regulation for top predators. }}
    \label{Sfig: GBR 5 eigenparameters}
\end{figure}

\begin{figure}[H]
    \centering
    \includegraphics[width=0.95\textwidth]{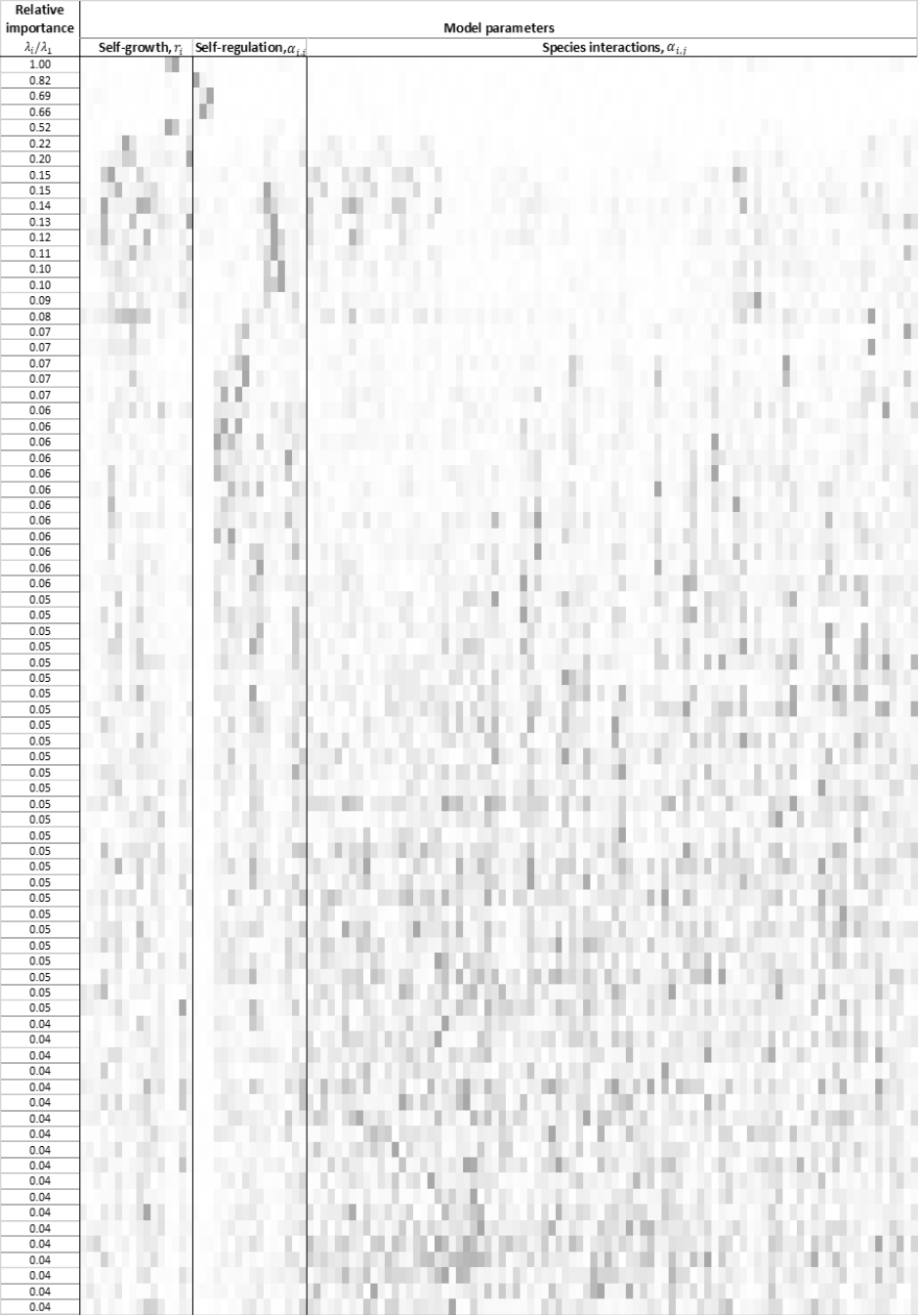}
    \caption{{\edits \textbf{Eighty most tightly constrained parameter combinations for the Great Barrier Reef ecosystem network. } The eigenvector values for the first 80 eigenparameters, shaded such that darker colours indicate a greater contribution of the parameter to the eigenparameter. Each row represents an eigenparameter (ordered from most sensitive to least) and each column represents a model parameter (grouped by type). Note that beyond the first five eigenparameters, there are no clearly interpretable trends. }}
    \label{Sfig: GBR 100 eigenparameters}
\end{figure}

\nolinenumbers
\section*{Acknowledgments}
The authors thank Chris Baker, Cailan Jeynes-Smith, Brodie Lawson and Robert Salomone for helpful discussions during this research. Computational resources were provided by the eResearch Office, Queensland University of Technology.

\nolinenumbers
\bibliographystyle{vancouver}


\newpage
\begin{center}
   
    \vspace*{1cm}
    \Large{
    \textbf{S1 File. Unlocking ensemble ecosystem modelling for large and complex networks}}
       
    \large{Sarah A. Vollert$^{a,b,*}$, Christopher Drovandi$^{a,b}$, \& Matthew P. Adams$^{a,b,c}$} \\

    \normalsize
    $^a$Centre for Data Science, Queensland University of Technology, Brisbane, Australia \\

    $^b$School of Mathematical Sciences, Queensland University of Technology, Brisbane, Australia\\

    $^c$School of Chemical Engineering, The University of Queensland, St Lucia, Australia\\

    $^*$Corresponding author. E-mail: sarah.vollert@hdr.qut.edu.au
  
   \end{center}
\appendix
\subsection*{Additional details of the SMC-EEM method}

\begingroup
\renewcommand{\section}[2]{}%
\renewcommand\thefigure{\thesection.\arabic{figure}}    
\setcounter{figure}{0} 
\renewcommand{\thealgocf}{S\arabic{algocf}}
\renewcommand\thetable{S\arabic{table}}    
\setcounter{table}{0} 
\setcounter{algocf}{0}

Algorithm 2 of the manuscript provides an overview of the SMC-EEM method. Here, we provide additional details for implementing the method (Algorithms \ref{Alg:SMC-based EEM} and \ref{Alg:EEM-MCMC step}), adapted from Drovandi and Pettitt's  implementation of SMC-ABC \cite{drovandi_2011_ABC}. Algorithm \ref{Alg:SMC-based EEM} describes the SMC-ABC algorithm including reweighting, resampling and moving the particles. Algorithm \ref{Alg:EEM-MCMC step} details the MCMC-ABC process used within the move step of Algorithm \ref{Alg:SMC-based EEM}. 

A key difference between the algorithm presented in the manuscript (Algorithm 2) and this more detailed version (Algorithms \ref{Alg:SMC-based EEM} and \ref{Alg:EEM-MCMC step}) is the inclusion of a bijective transform to the parameters to ensure that the reparameterisation is free from any constraints on the original parameters \cite{drovandi_2011_ABC}, such as the bounds of a uniform prior distribution (e.g.\ see Section S.4.2 of \citep{Vollert_2022}). The bijective transform ensures that any MCMC-ABC proposal will respect the constraints on the original parameters. The SMC-ABC samples for the original parameterisation can be easily obtained by applying the inverse transform to each SMC-ABC sample at the end of the algorithm. The implied prior distribution in the transformed space can be calculated via a transformation of random variables.

\begin{algorithm}[H]
\footnotesize
    \BlankLine
    \textbf{INITIALISE} \\
    \BlankLine
    Define the discrepancy function, $\rho(\bm{\theta})$ \\
    Specify the prior distribution, $\pi(\bm{\theta})$ \\
    Select the tuning variables, including: \\ 
    \qquad The number of particles to be sampled, $M$ \\ 
    \qquad The percentage of particles retained in each sequential step, $a$ \\
    \qquad The desired probability of particles unmoved during MCMC-ABC, $c$ \\
    \qquad The number of trial MCMC-ABC steps to gauge acceptance rate, $n_{\mathrm{MCMC}}$ \\ 
    \BlankLine
    Generate a sample of $M$ particles ($\{\bm{\theta}_i\}_{i=1}^{M}$) from the prior distribution, $\pi(\bm{\theta})$ \\    
    \BlankLine
    \BlankLine
    \textbf{REWEIGHT} \\
    \BlankLine
    Evaluate the discrepancy for all particles $\bm{\rho} = \{\rho(\bm{\theta}_{i})\}_{i=1}^{M}$ \\
    Sort the particles $\bm{\theta}$ in ascending order of their corresponding discrepancy $\bm{\rho}$\\
    Set the discrepancy threshold $\epsilon_t$ based on the number of particles to be retained $n_{\mathrm{keep}}=\mathrm{floor}(a\times M)$ \\
    \BlankLine
    \BlankLine
    \While{there are infeasible or unstable models in the ensemble, $\max(\bm{\rho}) > 0$} {
    \BlankLine
    \textbf{RESAMPLE} \\
    \BlankLine
    Transform current values of $\bm{\theta}$ to $\tilde{\bm{\theta}}$ such that the parameter-space is less restricted \\
    Duplicate retained particle values based on discrepancy $\bm{\rho}$ to replace those with $\rho_i>\epsilon_t$ \\
    Calculate the sample covariance matrix, $\Sigma = cov(\{\bm{\theta}_{i}\}_{i=1}^{n_{\mathrm{keep}}})$ \\
    \BlankLine
    \BlankLine
    \textbf{MOVE} \\
    \BlankLine
    \For{each of the $n_{\mathrm{MCMC}}$ trial MCMC-ABC steps}{
        Move the particles using MCMC-ABC (Algorithm \ref{Alg:EEM-MCMC step})
        \BlankLine
    }
    Estimate the {\edits MCMC-ABC acceptance rate for iteration $t$}, $a_t$ \\
    Determine the number of MCMC-ABC iterations to perform, $R_t = \lceil \log (c)/\log(1-a_t) \rceil$ {\edits and update $n_{MCMC}=R_t/2$} \\
    \For{each of the remaining MCMC-ABC steps, $R_t-n_{\mathrm{MCMC}}$}{
        Move the particles using MCMC-ABC (Algorithm \ref{Alg:EEM-MCMC step})
        \BlankLine
    }
    \BlankLine
    \BlankLine
    \textbf{REWEIGHT} \\
    \BlankLine
    Sort the particles $\bm{\theta}$ in ascending order of their corresponding discrepancy $\bm{\rho}$\\
    Set the discrepancy threshold based on the number of particles to be retained, $\epsilon_t = \rho(\bm{\theta}_{n_{\mathrm{keep}}})$\\
    \If{we are dropping feasible and stable particles, $\rho(\bm{\theta}_i) =0$ where $i > n_{\mathrm{keep}}$}{
    Adjust $n_{\mathrm{keep}}$ to retain all feasible and stable particles where $\rho(\bm{\theta}_i) =0$
    }
    Ensure all $\tilde{\bm{\theta}}$ are transformed back to  $\bm{\theta}$ \\
    \BlankLine
    }
    \caption{SMC-EEM algorithm used for sampling the ensemble of feasible and stable ecosystem models.}
\label{Alg:SMC-based EEM}
\end{algorithm}

\begin{algorithm}[H]
\footnotesize
        \For{each particle $i$ in $\{\bm{\theta}_{i}\}_{i=n_{\mathrm{keep}}}^{M}$}{
        Propose a new set of parameter values  ${\tilde{\bm{\theta}_i}}^*$ using a multivariate normal proposal distribution, ${\tilde{\bm{\theta}_i}}^* \sim N(\tilde{\bm{\theta}_i},\Sigma)$ \\        
        Calculate the prior probability ($\pi(\tilde{\bm{\theta}})$ for the transform space) of the current and proposed parameter values ($\tilde{\bm{\theta}_i}$ and ${\tilde{\bm{\theta}_i}}^*$) \\
        Transform the current and proposed parameter values ($\tilde{\bm{\theta}_i}$ and ${\tilde{\bm{\theta}_i}}^*$) in terms of $\bm{\theta}$ \\
        Evaluate the discrepancy $\rho(\bm{\theta}_{i}^*)$  \\
        Accept or reject a particle based on a Metropolis-Hastings acceptance probability $\alpha = \min \left(1, \pi(\tilde{\bm{\theta}}_i^*)/\pi(\tilde{\bm{\theta}}_i) \right)$, if within the discrepancy threshold, $\rho(\bm{\theta}_{i}^*) \leq \epsilon_t$ 
        }
    \caption{MCMC-ABC algorithm used within the SMC-EEM (Algorithm \ref{Alg:SMC-based EEM}) }
\label{Alg:EEM-MCMC step}
\end{algorithm}

{\edits Practitioners should note that within this algorithm there are three tuning parameters to be selected, whose values can have substantial impact on computation time and posterior samples if poorly chosen. These tuning parameters, their potential effects, and our suggested values are outlined in Table \ref{tab:tuning_parameters}. 

\begin{table}[H]
    \centering \edits
    \begin{tabular}{m{0.17\textwidth}|m{0.30\textwidth}|m{0.37\textwidth}|m{0.07\textwidth}}
        \textbf{Parameter} & \textbf{Tuning effect} & \textbf{Ensemble quality / efficiency tradeoff} &\textbf{Value used}\\ \hline
        $\bm{a}$: the percentage of particles retained during each iteration & Retaining more particles means that some higher discrepancy particles are retained and the space around these parameter sets are explored via MCMC. & Higher values of $a$ leads to quality ensembles (more parameter space explored), but slower computation times (more iterations needed to remove high-discrepancy particles). & 40\%\\ \hline
        $\bm{c}$: the desired probability of particles unmoved during MCMC-ABC & Lower probabilities of unmoved particles decreases the number of duplicate particles  retained during each iteration. & Higher values of $c$ leads to poorer quality ensembles (more duplicate parameter sets), but faster computation times (less time needed to remove duplicates). & 1\% \\ \hline
        $\bm{n_{MCMC}}$: the number of trial MCMC-ABC steps used to estimate the MCMC-ABC acceptance rate for each iteration & Once the acceptance rate is estimated, this is used to calculate how many MCMC-ABC steps will be needed to obtain the desired probability of duplicate parameters ($c$). & Higher values of $n_{MCMC}$ leads to quality ensembles (accurate estimates of acceptance rate can avoid excess duplicate particles), but slower computation times (more MCMC steps than necessary used to estimate the acceptance rate). & 10\\
    \end{tabular}
    \caption{Tuning parameters of the SMC-ABC algorithm. }
    \label{tab:tuning_parameters}
\end{table}

These parameters can be tuned to balance the trade-off between representative samples and computational efficiency. However, we suggest using the values we have recommended here and assessing the reproducibility of the posterior sample via multiple independent algorithm runs. We note that our tuning parameter choices could be considered conservative, as they favour high sample diversity over a faster and more aggressive algorithm. 

}

\endgroup

\subsection*{S1 File References}
\begingroup
\renewcommand{\section}[2]{}%
\bibliographystyle{vancouver}


\endgroup
\end{document}